\documentclass[11pt, a4paper]{article}
\usepackage{graphicx}
\usepackage{epsfig}
\usepackage{amsmath}
\usepackage{amsfonts}
\usepackage{amssymb}
\usepackage{epstopdf}
\usepackage{amsthm}
\usepackage[usenames]{color}
\usepackage{array}
\usepackage{xcolor}
\usepackage{float}
\usepackage{subfigure}
\usepackage{arydshln}

\usepackage[
      colorlinks=true,
      linkcolor=blue,
      urlcolor=blue,
      filecolor=blue,
      citecolor=red,
      pdfstartview=FitV,
      pdftitle={},
      pdfauthor={},
      pdfsubject={},
      pdfkeywords={},
      pdfpagemode=None,
      bookmarksopen=true
]{hyperref}

\newcommand{\mysection}[1]{\section{#1}\setcounter{equation}{0}}

\def\cA{{\cal A}}
\def\cE{{\cal E}}
\def\cR{{\cal R}}
\def\cS{{\cal S}}
\def\cT{{\cal T}}
\def\fg{{\mathfrak g}}

\def\be{\begin{equation}}
\def\ee{\end{equation}}
\def\beq{\begin{eqnarray}}
\def\eeq{\end{eqnarray}}

\begin{document}

\pagestyle{myheadings}
\markboth{\textsc{\small }}{
\textsc{\small Scanning the parameter space of collapsing rotating thin shells}} \addtolength{\headsep}{4pt}


\begin{centering}
  \textbf{\Large{Scanning the parameter space of collapsing rotating thin shells}}

\vspace{1.5cm}

  \large{Jorge V. Rocha$^{1,\sharp}$ and Raphael Santarelli$^{2,\natural}$}

\vspace{1cm}

\begin{minipage}{.9\textwidth}
  \small \it
    \begin{center}
$^1$
Departament de F\'isica Qu\`antica i Astrof\'isica, Institut de Ci\`encies del Cosmos (ICCUB),\\
Universitat de Barcelona, Mart\'i i Franqu\`es 1, E-08028 Barcelona, Spain
\\ $ \, $ \\

$^2$
Departamento de F\'isica, Universidade Federal de S\~ao Carlos,\\ P. O. Box 676, 13565-905, S\~ao Carlos, S\~ao Paulo, Brazil
\\ $ \, $    \end{center}
\end{minipage}

\end{centering}

\vspace{0.5cm}

\begin{abstract}
We present results of a comprehensive study of collapsing and bouncing thin shells with rotation, framing it in the context of the weak cosmic censorship conjecture. The analysis is based on a formalism developed specifically for higher odd dimensions that is able to describe the dynamics of collapsing rotating shells exactly.
We analise and classify a plethora of shell trajectories in asymptotically flat spacetimes. The parameters varied include the shell's mass and angular momentum, its radial velocity at infinity, the (linear) equation-of-state parameter and the spacetime dimensionality.
We find that plunges of rotating shells into black holes never produce naked singularities, as long as the matter shell obeys the weak energy condition, and so respect cosmic censorship. This applies to collapses of dust shells starting from rest or with a finite velocity at infinity. Not even shells with a negative isotropic pressure component (i.e., tension) lead to the formation of naked singularities, as long as the weak energy condition is satisfied.
Endowing the shells with a positive isotropic pressure component allows the existence of bouncing trajectories satisfying the dominant energy condition and fully contained outside rotating black holes. Otherwise any turning point occurs always inside the horizon.
These results are based on strong numerical evidence from scans of numerous sections in the large parameter space available to these collapsing shells.
The generalisation of the radial equation of motion to a polytropic equation-of-state for the matter shell is also included in an appendix.
\end{abstract}

\vfill

\noindent
\rule{6.2cm}{0.2mm}
\\
\noindent
{\small
$^\sharp$ \texttt{jvrocha@icc.ub.edu}\\
$^\natural$ \texttt{santarelli@df.ufscar.br}
}

\thispagestyle{empty}
\newpage


\mysection{Introduction}

While the study of spherical gravitational collapse of stars leading to black holes (BHs) has a long history dating back to the work of Oppenheimer and Snyder~\cite{Oppenheimer:1939ue}, full investigations of collapsing matter carrying angular momentum only became possible with the advent of numerical relativity in the '80s~\cite{Nakamura:1981, Stark:1985da}.
Of course, the reason for this historic delay is that the introduction of rotation in three spatial dimensions typically breaks spherical symmetry, thus increasing the complexity of the problem.

Besides uncovering the fate of realistic stars, such studies have an important bearing on the weak cosmic censorship conjecture~\cite{Penrose:1969pc}, which is still an issue of intense debate.
In essence, this asserts that any curvature singularities forming from generic collapse of physically reasonable matter remain hidden inside black hole horizons. In other words, the conjecture forbids the development of naked singularities, preventing quantum gravity spacetime regions to be accessed by asymptotic observers.

Axisymmetric simulations of stellar collapse have shown that whether or not a BH forms depends strongly on the amount of angular momentum~\cite{Shibata:2000gt}. The endpoint of a full collapse process is expected to be a stationary spacetime, which in four dimensions ---and assuming departures from vacuum are negligible--- must belong to the Kerr family of solutions~\cite{Kerr:1963}, parametrised by mass $M$ and angular momentum $J$. When $J/M^2 \leq1$ there is a BH horizon covering the curvature singularity, but if $J/M^2 >1$ one obtains a naked singularity. It is then interesting to ask what happens when an initial configuration with $J/M^2 >1$ undergoes gravitational collapse. This problem was addressed in Refs.~\cite{Abrahams:1994ge, Giacomazzo:2011cv}, where it was found that in potentially dangerous over-spinning cases, naked singularities were never formed. Instead, other equilibrium configurations arose after some complicated transient dynamics.
Similarly, studies of grazing high-energy black hole collisions~\cite{Sperhake:2009jz} indicate that any excessive angular momentum is radiated away until a sub-extremal ($J/M^2 \leq1$) regime is reached and the merger always produces a final Kerr black hole.

There is a long history of assessments of cosmic censorship based on stress-testing the stability of vacuum black hole horizons under the absorption of test particles. The first such study was the seminal work of Wald~\cite{Wald:1974}, showing that 4D extremally rotating black holes cannot be over-spun with test particles. Those with sufficiently large angular momentum to raise the joint BH+particle system above the Kerr bound are simply scattered by the black hole, leaving the horizon untouched. This picture has been extended to higher dimensions~\cite{BouhmadiLopez:2010vc} with similar results; Refs.~\cite{Rocha:2014gza,Rocha:2014jma} specifically concern spacetimes with equal angular momenta, which will be the focus of the present paper. 
Test fields interacting with extremal black holes were also shown to comply with weak cosmic censorship~\cite{Natario:2016bay}. The consideration of near-extremal BHs ---as opposed to exactly extremal--- opened a narrow window of opportunity to overspin 4D black holes with point particles, by neglecting back-reaction effects~\cite{Jacobson:2009kt}. However, a proper account of the self-force was argued to restore the validity of the conjecture~\cite{Barausse:2010ka}. Very recently, this was confirmed to be the case, by incorporating crucial self-force effects of second order in the angular momentum of a body falling into the black hole~\cite{Sorce:2017dst}.
It should also be mentioned that some notable violations of cosmic censorship do occur in higher-dimensional general relativity~\cite{Lehner:2010pn,Figueras:2015hkb,Figueras:2017zwa}, but these are of a completely different nature: they are a consequence of instabilities afflicting extended BH horizons. Therefore, such objects are not expected to result generically from gravitational collapse.

The topic of gravitational collapse in the presence of rotation has hardly been explored with analytic methods.
Some early partial results were obtained by employing approximations, such as the assumptions of adiabatic collapse, or slow rotation~\cite{Cohen:1968, Lindblom:1974bq}. Alternatively, Ref.~\cite{Wagoner:1965} made progress by focusing attention on the dynamics only in a neighbourhood of the equatorial plane.
Clearly, the restriction to two spatial dimensions bypasses the main technical hurdle, since it allows rotating configurations while avoiding any dependence on angular coordinates. This was used to study gravitational collapse in $(2+1)$-dimensional spacetimes for the case of thin rings of matter~\cite{Crisostomo:2003xz,Mann:2008rx} and for inhomogeneous disks of dust~\cite{Vaz:2008uv}. In these contexts, naked singularities never arise from collapses with rotation, as long as matter obeys the weak energy condition, i.e., weak cosmic censorship is observed.

Nevertheless, it is in fact possible to tackle the problem of gravitational collapse with rotation ---without restricting to lower dimensions--- in a fairly simple way~\cite{Delsate:2014iia}. The idea relies on the consideration of equal angular momenta (EAM) spacetimes. This possibility arises in higher odd dimensions ($D=5,7,9,...$), allowing rotating geometries to depend on a single coordinate --- for this reason they are referred to as cohomogeneity-1 geometries. This results from an enhanced symmetry, in practice making them resemble spherically symmetric spacetimes. The formalism for rotating thin matter shells in 5D was developed in Ref.~\cite{Delsate:2014iia}, and more recently has been extended to higher odd dimensions~\cite{Rocha:2015tda}. A similar idea was first used in Ref.~\cite{Bizon:2005cp} to study critical collapse in 5D vacuum gravity, although that investigation was restricted to non-rotating spacetimes.

The goal of the present paper is to perform a comprehensive study of exact collapses of rotating thin shells. Within the framework of Refs.~\cite{Delsate:2014iia,Rocha:2015tda} we can assess the effect of rotation on gravitational collapse and consequently on cosmic censorship. Since the full evolution of the thin shell spacetime is obtained by a simple integration of a radial effective potential that is known {\em exactly}, we can easily scan a large parameter space. The parameters we vary include the proper mass of the shell, the angular momentum, the radial velocity of the shell at infinity, a linear equation-of-state parameter and the spacetime dimensionality.

The output of this parameter scan is presented in the final section of the paper, alongside with ample discussion of the results. Nonetheless, it is convenient to highlight here our main findings:
\begin{itemize}
\item We observe no violation of the weak cosmic censorship conjecture. In the context of equal angular momenta spacetimes, collapses of thin shells onto rotating black holes can destroy the horizon only if the matter shell violates the weak energy condition. This is also true for the less-restrictive null energy condition\footnote{Here we are considering the energy conditions from the viewpoint of the shell's worldsheet. One can also assess the energy conditions as derived from a stress-energy tensor on the full spacetime, but being localized on the timelike hypersurface of the shell. The two descriptions are related: the weak energy condition on the shell is equivalent to the null energy condition on the full spacetime~\cite{Natario:2017szw}.}.
\item Full plunges of rotating dust shells ---for which the isotropic component of the pressure vanishes, although the rotation induces a nonzero anisotropic pressure--- always violate the dominant energy condition before hitting the singularity.
\item Still considering rotating dust shells, the dominant energy condition can be satisfied if there is a turning point, which can either be inside the horizon or else if there is no horizon at all. The former case corresponds to a two-world orbit in which the shell crosses a black hole horizon and later emerges from a white hole into a different universe. The latter case describes a shell bouncing off a naked singularity.
\item The consideration of a nonzero isotropic pressure component opens up more possibilities (but the shell must approach the speed of light at infinity). In this case there is an interesting competition between centrifugal forces and pressure, which allows bounces with turning points outside a black hole horizon, while obeying the dominant energy condition.
\end{itemize}

Our approach offers some advantages over previous work concerning the gravitational collapse of rotating shells. Contrary to Ref.~\cite{Lindblom:1974bq} (for a textbook exposition, see section 3.10 of~\cite{Poisson:2004}), we can follow the shell's entire trajectory exactly. The point is that in descriptions adopting a small rotation approximation the solutions cease to be valid near the ergosphere, and therefore they cannot say anything about full collapse to a black hole (or a naked singularity).
Moreover, compared to Refs.~\cite{Crisostomo:2003xz,Mann:2008rx} which focused on the shell's trajectory, we also take particular care in checking whether energy conditions are satisfied during the orbit, since this plays an important role in the formulation of cosmic censorship.

The framework adopted to construct rotating thin shell spacetimes also faces a few limitations.
The most obvious one is that the consideration of cohomogeneity-1 rotating geometries restricts us to odd spacetime dimensions.
Secondly, the matching of two such backgrounds to obtain the thin shell is performed in the simplest possible way, guaranteeing that the angular symmetry group ---which turns out to be $U\left(\frac{D-1}{2}\right)$--- is preserved by the global spacetime. This limits our ability to construct rotating thin shells to cases in which the interior region already has a black hole or a naked singularity; i.e., it cannot be flat. Thus, the scenarios considered correspond to collapses (or bounces) {\em onto} black holes or naked singularities and, therefore, are not suitable for studies of critical collapse.\footnote{This does not mean that more involved matching surfaces cannot avoid this feature, but then we would encounter the same difficulties found in the construction of 4D rotating thin shells.}
Finally, since both the exterior and interior of the shell are taken to be stationary ---namely Myers-Perry solutions~\cite{Myers:1986un} with all spin parameters set equal--- clearly there will be no gravitational radiation, even though the shell simultaneously contracts/expands and rotates. Hence, these are curiously special spacetimes. Nevertheless, from the point of view of testing cosmic censorship this is precisely the most dangerous scenario, since gravitational radiation is typically much more efficient in dissipating angular momentum than energy~\cite{Shibata:2008rq,Pollney:2009yz}.

The remainder of the paper is organised as follows. We start by reviewing the framework employed to construct rotating thin shell spacetimes in Section~\ref{sec:matching}. Then we analyse energy conditions in Section~\ref{sec:energyconds}. In Section~\ref{sec:potential} we study general properties of the radial dynamics of rotating shells. Finally, the scan of the parameter space of collapsing rotating thin shells is performed in Section~\ref{sec:scan}, where we also include the discussion of the results.
We relegate to Appendix~\ref{sec:boundsEC} some general bounds obtained from the weak and dominant energy conditions. In Appendix~\ref{sec:poly} we present the generalisation of the radial equation of motion to shells with polytropic equations-of-state.

\mysection{Cohomogeneity-1 thin shell spacetimes
\label{sec:matching}}

This section reviews material covered in Refs.~\cite{Delsate:2014iia, Rocha:2015tda}, where more details can be found. Therefore, we will only summarise the main points, which will also serve to fix important notation.
Our focus here is on asymptotically flat spacetimes, but note that the analysis can be easily extended to include a nonvanishing cosmological constant.
We will work in geometrised units, for which the gravitational constant and the speed of light are set to unity, $G=c=1$.

\bigskip
We take the interior and exterior spacetimes (indicated with subscripts $-$ and $+$ on all associated quantities, respectively) to be EAM Myers-Perry solutions in $D=2N+3$ dimensions, with $N$ an integer. This family of stationary geometries has enhanced symmetry and their line element depends essentially only on a radial coordinate~\cite{Frolov:2002xf,Kunduri:2006qa}:
\be
  ds_\pm^2 = g_{\mu\nu} dy^\mu dy^\nu  =  - f_\pm(r)^2 dt^2 + g_\pm(r)^2 dr^2 
 + h_\pm(r)^2 \left[ d\psi + A_a dx^a - \Omega_\pm(r) dt \right]^2
 + r^2 \widehat{g}_{ab} dx^a dx^b \,,
\label{eq:metric}
\ee
where
\begin{flalign}
&  g_\pm(r)^2 = \left( 1 - \frac{2M_\pm}{r^{2N}} + \frac{2M_\pm a_\pm^2}{r^{2N+2}} \right)^{-1}\,, 
\label{eq:metricfuncs1}\\
&  h_\pm(r)^2=r^2\left(1+\frac{2M_\pm a_\pm^2}{r^{2N+2}} \right)\,, \qquad \Omega_\pm(r)=\frac{2M_\pm a_\pm}{r^{2N} h_\pm(r)^2}\,, \qquad  f_\pm(r)=\frac{r}{g_\pm(r) h_\pm(r)}\,.
\label{eq:metricfuncs2}
\end{flalign}
Here, $M_\pm$ and $a_\pm$ are the mass and spin parameters, respectively.
This form of writing the metric relies on the description of the constant $t$- and $r$-slices ---which are topologically $(2N+1)$-spheres--- as a $S^1$ bundle over the complex projective space $CP^N$.
The $x^a$ are the $2N$ coordinates on $CP^N$, which is endowed with the standard Fubini-Study line element $\widehat{g}_{ab}dx^a dx^b$, and $A=A_a dx^a$ is the associated K\"ahler potential. Explicit expressions for $\widehat{g}_{ab}$ and $A_a$ can be obtained iteratively in $N$~\cite{Hoxha:2000jf,Dias:2010eu} but we shall not require them. The coordinates $y^\mu$ cover the whole manifold and run over $\{t,r,\psi,x^a\}$. The $S^1$ fiber is parametrised by the angular coordinate $\psi$, with periodicity $2\pi$.
The metrics~\eqref{eq:metric} are vacuum solutions of the Einstein equations.
The largest real root of $g_\pm^{-2}$ indicates an event horizon whose spatial sections have the geometry of a homogeneously squashed $(2N+1)$-sphere. When the parameters $M_\pm$ and $a_\pm$ are such that the function $g_\pm^{-2}$ does not have zeroes, the associated spacetime corresponds to a naked singularity, with the curvature diverging at $r=0$.

\bigskip
The next step is to match two spacetimes with line elements of the form~\eqref{eq:metric} across a timelike hypersurface $\Sigma$ defined by the parametric equations $t=\cT(\tau)$ and $r=\cR(\tau)$, where $\tau$ is the proper time of an observer comoving with the hypersurface.

The first Darmois-Israel junction condition~\cite{Israel:1966rt,Darmois} demands the continuity of the metric across the thin shell, so that the induced metrics match, $\fg_{ij}^{(+)} = \fg_{ij}^{(-)} \equiv \fg_{ij}$. This has two immediate consequences: since the parameter $\tau$ is taken to be the proper time, this implies a relation between $\dot{\cT}$ and $\dot{\cR}$,
\be
f(\cR)^2 \dot{\cT}^2 - g(\cR)^2 \dot{\cR}^2=1\,,
\label{eq:TRtau}
\ee
where an overdot stands for $d/d\tau$.
In addition, the first junction condition imposes a rigid relationship between the parameters of the interior and exterior geometries,
\be
h_+(\cR)=h_-(\cR)\equiv h(\cR)   \qquad  \Rightarrow  \qquad   M_+a_+^2 = M_-a_-^2\,.
\label{eq:Ma2}
\ee
This implies that a such a rotating cohomogeneity-1 exterior (with $M_+a_+\neq0$) cannot be continuously joined with a flat interior ($M_-=0$) along a constant-$r$ surface.

\bigskip
The second junction condition fixes the form of the surface stress-energy tensor $\cS_{ij}$ that sources any possible divergences on $\Sigma$.
\be
\cS_{ij} = -\frac{1}{8\pi \kappa_N} \left( [[k_{ij}]] - \fg_{ij} [[k]] \right)  \,,
\label{eq:junction2}
\ee
where $k_{ij}$ represents the extrinsic curvature and $k=\fg^{ij}k_{ij}$ is its trace. Here we have introduced
\be
  [[C_{ij\dots}]]\equiv C_{ij\dots}^{(+)}-C_{ij\dots}^{(-)}
\ee
as a short-hand notation for the jump of any given tensorial quantity $C_{ij\dots}$ across $\Sigma$. We included a dimension-dependent numerical prefactor on the right hand side, $\kappa_N$, which will be fixed in Section~\ref{sec:potential}. For $N=1/2$, corresponding formally to standard four-dimensional gravity, we should recover $\kappa_{1/2}=1$.

One might be tempted to assume $\dot{\cT}>0$. After all, this must be the case when a timelike shell is outside the black hole event horizon or inside the Cauchy horizon. Note however, that between the event horizon and the Cauchy horizon (where $g(\cR)^2<0$) the derivative $\dot{\cT}$ can in fact change sign.\footnote{From the point of view of the maximal analytic extension, the sign of $\dot{\cT}$ dictates which of the two allowed trajectories the shell follows in the domain between the event horizon and the Cauchy horizon~\cite{Gao:2008jy}.} As we will see, this overall sign does not affect neither the conservation equations nor the shell's equation of motion.

The various components of the extrinsic curvature were computed in~\cite{Rocha:2015tda}.
The form of the stress-energy tensor $\cS_{ij}$ is then dictated by the second junction condition~\eqref{eq:junction2},
\be
\cS_{ij} = (\rho+P)u_iu_j + P\fg_{ij} + 2\varphi\, u_{(i}\xi_{j)} + \Delta P\, \cR^2 \widehat{g}_{ab} \frac{dx^a}{dy^i} \frac{dx^b}{dy^j}\,,
\label{eq:Sij}
\ee
where coordinates $y^i$ run over $\{\tau,\psi,x^a\}$.
This stress-energy tensor describes an imperfect fluid.
Here, $u=u^i \partial_i=\partial_\tau$ is the normalised fluid velocity (assumed to be corotating with the shell), and $\xi=\xi^i \partial_i=h^{-1}\partial_\psi$ is a unit vector aligned with the $S^1$ fiber (the direction that effectively incorporates the rotation of the spacetime).
The quantity $\varphi$ is commonly referred to as heat flow, and it can be thought of as an intrinsic momentum of the fluid, while $\Delta P$ denotes the pressure anisotropy.

More explicitly, the components of the stress-energy tensor are given by the following expressions:
\begin{flalign}
& \rho= - \frac{[[\beta(\cR)]]}{8\pi\kappa_N\,\cR^{2N+1}} \frac{d}{d\cR}\big[\cR^{2N}h(\cR)\big]\,, \label{eq:rho}\\
& P= \frac{h(\cR)}{8\pi\kappa_N\,\cR^{2N+1}} \frac{d}{d\cR}\big[\cR^{2N}[[\beta(\cR)]]\big]\,, \label{eq:P}\\
& \varphi= - \frac{h(\cR)^2}{16\pi\kappa_N\,\cR} [[\Omega'(\cR)]]\,, \label{eq:varphi}\\
& \Delta P= \frac{[[\beta(\cR)]]}{8\pi\kappa_N} \frac{d}{d\cR}\left[\frac{h(\cR)}{\cR}\right]\,. \label{eq:DeltaP}
\end{flalign}
where for convenience we defined another quantity:
\be
\beta_\pm \equiv {\rm sign}(\dot{\cT}) f_\pm \sqrt{1+g_\pm^2\dot{\cR}^2}\,.
\label{eq:beta}
\ee

From these expressions it follows immediately that in the absence of rotation both $\varphi$ and $\Delta P$ vanish, and so we recover a perfect fluid. Thus, the heat flow and the pressure anisotropy are induced by the shell's rotation.

One can verify that such a stress-energy tensor is covariantly conserved. The conservation equations, $\nabla^i\cS_{ij}=0$, reduce to
\begin{flalign}
& - \frac{d}{d\cR}\left(h \cR^{2N}\rho\right) = \left(2N \frac{h}{\cR}+h' \right)\cR^{2N}P + 2N h \cR^{2N-1}\Delta P\,,
\label{eq:rhoPbalance} \\
& \frac{d}{d\tau}\left[\varphi\,\cR^{2N}h(\cR)^2\right]=0\,.
\label{eq:AMconserv}
\end{flalign}
The latter equation is automatically satisfied with $\varphi$ given by~\eqref{eq:varphi} and taking into account the definitions of the metric functions~\eqref{eq:metricfuncs2}. Similarly, Eq.~\eqref{eq:rhoPbalance} is obeyed with the energy density, pressure and pressure anisotropy prescribed by expressions~(\ref{eq:rho}), (\ref{eq:P}) and (\ref{eq:DeltaP}), respectively.

Equation~\eqref{eq:AMconserv} simply expresses the conservation of the shell's angular momentum during evolution. Indeed, the quantity within brackets is proportional to the jump in the angular momentum across the shell,
\be
\varphi\,\cR^{2N}h(\cR)^2 = -\frac{[[Ma]]}{4\pi\kappa_N}\,.
\ee
As for Eq.~\eqref{eq:rhoPbalance}, it affords a clear interpretation: the (intrinsic) energy gained by the shell as it shrinks is accounted for by the work done by the pressure components. The shell's surface area is proportional to $h(\cR) \cR^{2N}$ so a change in radius $d\cR$ implies a change in area equal to $\left(2N \frac{h}{\cR}+h' \right) \cR^{2N} d\cR$. This gives precisely the factor in front of the isotropic pressure component $P$. The factor appearing in front of the pressure anisotropy $\Delta P$ is instead $2N h \cR^{2N-1}$, because this component only acts on the $CP^N$ coordinates $x^a$, i.e., it is not sensitive to changes in area along the $\psi$ direction.

\mysection{Energy conditions
\label{sec:energyconds}}

The standard energy conditions are generally specified as inequalities imposed on the stress-energy tensor, when contracted with arbitrary timelike or null vectors~\cite{Wald:1984rg}. In practice, it is useful to translate this into explicit constraints on the stress-energy components. When applied to a perfect fluid, this yields very simple inequalities to be satisfied by the energy density and pressure. For the case of imperfect (viscous) fluids, such conditions have been worked out in Ref.~\cite{Kolassis:1988}. We are unaware of any explicit energy conditions in the literature concerning the sort of anisotropic fluids we consider in this work. Therefore, we shall derive them in this section.

The energy conditions are most conveniently expressed in terms of the eigenvalues of the stress-energy tensor~\eqref{eq:Sij}. These are obtained as the coefficients $\lambda_n$, with $n=0,\dots,2N+1$, such that
\be
\det[\cS_{ij}-\lambda_n \fg_{ij}]=0\,,
\ee
and they are given by~\cite{Rocha:2015tda}
\beq
\lambda_0 &=& \frac{P-\rho}{2} - \sqrt{\left(\frac{P+\rho}{2}\right)^2-\varphi^2}\,, \label{eq:lambda0}\\
\lambda_1 &=& \frac{P-\rho}{2} + \sqrt{\left(\frac{P+\rho}{2}\right)^2-\varphi^2}\,, \label{eq:lambda1}\\
\lambda_\alpha &=& P+\Delta P\,, \qquad \alpha=2,\dots,2N+1. 
\eeq
Some comments are in order:
\begin{itemize}
\item
Firstly, the $2N$ eigenvalues $\lambda_\alpha$ are degenerate and it can be checked that their associated eigenvectors are all spacelike.
\item
One needs $(P+\rho)^2-4\varphi^2\geq0$ in order to have {\em real} eigenvalues $\lambda_0$ and $\lambda_1$ (as well as the corresponding eigenvectors). Otherwise, there are only $2N$ spacelike eigenvectors and the stress-energy tensor is of type IV. In this case, it cannot even satisfy the weak energy condition~\cite{Kuchar:1990vy}.
\item
In the limiting case $(P+\rho)^2-4\varphi^2=0$ there is ---in addition to the $2N$ spacelike eigenvectors associated with the eigenvalues $\lambda_\alpha$--- a double null eigenvector. This yields a type II stress-energy tensor.
\item
When $(P+\rho)^2-4\varphi^2>0$ there is a total of $2N+1$ spacelike eigenvectors and $1$ timelike eigenvector. The stress-energy tensor is of type I. The timelike eigenvector is the one associated with $\lambda_0$ as long as $\rho+P\geq0$, otherwise it is the one associated with $\lambda_1$. However, in the latter case it follows immediately that the condition $WEC_1$ below cannot be satisfied (with $\lambda_0$ and $\lambda_1$ interchanged). Therefore, only in the case $\rho+P\geq0$ can the weak energy condition be satisfied.
\end{itemize}

The weak energy condition (WEC) can now be formulated as a combination of the following inequalities~\cite{Wald:1984rg, Kuchar:1990vy}
\begin{flalign}
& WEC_0 \equiv -\lambda_0 \geq 0\,, \qquad\;\;
WEC_1 \equiv \lambda_1-\lambda_0 \geq 0\,, \qquad\;\;
WEC_\alpha \equiv \lambda_\alpha-\lambda_0 \geq 0\,, \nonumber\\
& 
WEC_t \equiv P+\rho\geq0\,, \qquad
WEC_r \equiv (P+\rho)^2-4\varphi^2\geq0 \,. 
\label{eq:WEC}
\end{flalign}
In the non-rotating case, the heat flow $\varphi$ and the pressure anisotropy $\Delta P$ both vanish and the weak energy condition reduces to the familiar relations: $\rho\geq0$ and $\rho+P\geq0$. In coordinate-invariant terms, the WEC requires that the double contraction of the stress-energy tensor with any timelike vector is nonnegative. 
The less restrictive null energy condition (NEC) possesses a similar coordinate-invariant definition but one considers instead null vectors, which amounts to simply omitting the first inequality in~\eqref{eq:WEC}.

The more physical dominant energy condition (DEC), which is typically obeyed by ordinary classical matter, imposes, in addition to~\eqref{eq:WEC}, the following inequalities:
\be
DEC_1 \equiv -\lambda_1-\lambda_0 \geq 0\,, \qquad
DEC_\alpha \equiv -\lambda_\alpha-\lambda_0 \geq 0\,.
\label{eq:DEC}
\ee
In the non-rotating limit we again retrieve the well-known relations for perfect fluids, namely $\rho\geq0$ and $-\rho\leq P\leq\rho$.

Some general bounds on parameters derived from these energy conditions are presented in Appendix~\ref{sec:boundsEC}.

\mysection{Radial dynamics and energy considerations
\label{sec:potential}}

Equations~(\ref{eq:rho}--\ref{eq:DeltaP}) determine the various components of the matter stress-energy tensor as a function of the shell's radial location and velocity. In order to close the system of equations one must specify an equation-of-state (EoS) relating the different components.
Here we adopt a linear EoS by taking the isotropic pressure to be proportional to the energy density,
\be
P=P(\rho)=w\rho\,.
\label{eq:linearEoS}
\ee
The extension of our study to a polytropic EoS is possible. The expressions become increasingly involved, so we relegate them to Appendix~\ref{sec:poly}.

Inserting Eqs.~\eqref{eq:rho} and \eqref{eq:P} into relation~\eqref{eq:linearEoS} we can easily integrate the equation to obtain
\be
[[\beta(\cR)]]= - \frac{m_0^{1+\frac{2N+1}{2N}w}}{\cR^{2N(1+w)}h(\cR)^w}\,,
\label{eq:deltabeta}
\ee
where $m_0$ is an integration constant with dimensions of mass.
We can now plug in the expressions for $\beta_\pm$, which depend on the shell's velocity as defined in \eqref{eq:beta}. The resulting equation can be cast in the standard form of an equation of motion for a classical particle moving in a one-dimensional radial potential,
\be
  \dot \cR^2 + V_{\rm eff}(\cR) =0\,,
\label{eq:motion}
\ee
where the effective potential $V_{\rm eff}$ is given explicitly by
\beq
V_{\rm eff}(\cR) &=& 1 + \frac{2Ma^2}{\cR^{2N+2}} - \frac{M_++M_-}{\cR^{2N}}  - \left(\frac{M_+-M_-}{m_0}\right)^2 \left(\frac{\cR^{2N}}{m_0}\right)^{\frac{2N+1}{N}w} \left(1+\frac{2Ma^2}{\cR^{2N+2}}\right)^{w-1} \nonumber\\
&& -\frac{1}{4} \left(\frac{m_0}{\cR^{2N}}\right)^{2+\frac{2N+1}{N}w} \left(1+\frac{2Ma^2}{\cR^{2N+2}}\right)^{1-w}.
\label{eq:potential}
\eeq
From its definition~\eqref{eq:motion}, classically allowed motion of the shell is restricted to radii satisfying $V_{eff}(\cR)\leq0$ and turning points occur when $V_{eff}(\cR)=0$.

Recall that $\dot{\cR} = d\cR/d\tau$ so, to obtain the velocity (squared) as seen by an asymptotic observer, one must convert from the shell's proper time $\tau$ to $\cT$. This is accomplished by using \eqref{eq:TRtau}, and the resulting radial potential becomes
\be
\widehat{V} \equiv -\left(\frac{d\cR}{d\cT}\right)^2 = \frac{\cR^2}{h(\cR)^2} \frac{g(\cR)^{-4} V_{eff}}{g(\cR)^{-2}-V_{eff}}\,.
\label{eq:VhatVeff}
\ee
In particular, when the shell approaches a horizon ---where $g^{-2}$ vanishes--- the asymptotic observer sees it slowing down to zero velocity, as expected. Also, a diverging $d\cR/d\tau$ when the shell is taken to infinity acquires a sound physical meaning: an asymptotic observer sees the shell approaching the speed of light, $d\cR/d\cT\to1$.

It is instructive to analyze simpler particular cases to gain some intuition. This is what we will do in the following. In the end of this section we return to the most general case.

\subsection{Dust shells}

We start by considering the case of rotating shells composed of dust. By `dust' we mean that the matter does not experience any isotropic pressure. In terms of the EoS parameter, this translates into $w=0$.

Now, in this case Eq.~\eqref{eq:deltabeta} evaluates to
\be
[[\beta(\cR)]] = \beta_+ - \beta_- = - \frac{m_0}{\cR^{2N}}\,.
\ee
The jump in the gravitational energy across the shell is then given by
\be
\Delta M=M_+-M_-=\frac{m_0}{2}(\beta_+ + \beta_-)\left(1+\frac{2Ma^2}{\cR^{2N+2}}\right)\,.
\label{eq:DeltaM}
\ee
This is easily derived by noting that $(\beta_+-\beta_-)(\beta_++\beta_-)=\beta_+^2-\beta_-^2=(g_+^{-2}-g_-^{-2})\cR^2/h(\cR)^2$  and it shows that $\Delta M\geq0$ if $m_0$ is positive.
We can equaly express this as
\be
\Delta M = E m_0\,,
\label{eq:MEm0}
\ee
where the energy per unit proper mass is defined by
\be
E \equiv \frac{(\beta_+ + \beta_-)}{2}\left(1+\frac{2Ma^2}{\cR^{2N+2}}\right)\,.
\ee
Although not apparent from this expression, Eq.~\eqref{eq:MEm0} shows that $E$ must be a constant of motion.

By squaring the relation~\eqref{eq:DeltaM} and inserting the definition of $\beta_\pm$, one can can express the total ADM mass in terms of the remaining quantities,
\be
M_+ = M_- +m_0\sqrt{1+\frac{2Ma^2}{\cR^{2N+2}}}\sqrt{1+\dot{\cR}^2+\frac{2Ma^2}{\cR^{2N+2}}-\frac{2M_-}{\cR^{2N}}} - \frac{m_0^2}{2\cR^{2N}}\left(1+\frac{2Ma^2}{\cR^{2N+2}}\right)\,.
\label{eq:DeltaMcontribs}
\ee
In the non-rotating case ($a=0$), one easily recognises the different contributions to the change $\Delta M$ in the total energy due to the shell: the square root term gives the relativistic kinetic energy of the shell (including rest mass), while the negative contribution proportional to $m_0^2$ represents the binding energy (see Ref.~\cite{Poisson:2004}, section 3.9). As expected, the presence of a black hole in the interior affects the total energy, yielding additional binding. We see that when rotation is included there are extra contributions to both the kinetic and the binding energy.

When the shell is taken to infinity, Eq.~\eqref{eq:DeltaMcontribs} reduces to $M_+\simeq M_-+m_0\sqrt{1+\dot{\cR}^2}$, which also shows that $\Delta M \geq m_0$ for these dust shells, i.e., $E\geq1$. Thus, we conclude that (for $w=0$)
\be
E = \left.\sqrt{1+\dot{\cR}^2}\right|_{\cR\to\infty}\,,
\label{eq:EdRdtau}
\ee
and having $E>1$ means that the shell has nonvanishing velocity at infinity.

The {\it intrinsic} energy of the shell, as seen by an observer comoving with the shell, can be computed as an integral over the volume of the shell,
\be
\cE=\int_0^{2\pi} d\psi \int d^{2N}x \sqrt{\det \gamma_{\alpha\beta}}\; n_i u_j \cS^{ij}\,,
\ee
where $\gamma_{\alpha\beta}$ is the induced metric on a $\tau=const.$ surface, $n^i=(1,0,\vec0)$ is a unit normal to this surface and $u^j=(\partial_\tau)^j$ is the unit timelike vector (the comoving observer velocity). Thus we obtain
\be
\cE= 2\pi {\rm Vol}(CP^N) h(\cR)\cR^{2N} \rho\,.
\ee
Inserting the result for the energy density, Eq.~\eqref{eq:rho}, one finds
\be
\cE= \frac{(2N+1)\cA_{2N+1}}{8\pi\kappa_N} m_0 \left[1+\frac{N}{2N+1}\frac{2Ma^2}{\cR^{2N+2}}\right]\,,
\ee
where ${\cal A}_{2N+1}$ is the area of a unit $2N+1$-dimensional sphere.
Thus, we conclude that the rotation contributes to the shell's intrinsic energy.
We can now fix the $N$-dependent factor so that the intrinsic energy $\cE$ precisely matches the rest mass $m_0$ at infinity:
\be
\kappa_N = \frac{(2N+1) \cA_{2N+1}}{8\pi} = \frac{(2N+1)\pi^N}{4\, \Gamma(N+1)} \,.
\ee
For $N=1/2$ (formally corresponding to $D=4$) we indeed get $\cE|_{\cR\to\infty}=m_0$ with $\kappa_{1/2}=1$.

Observe that, in the presence of rotation, the shell's intrinsic energy $\cE$ is {\em not} conserved, in contrast with $E$. In fact, we already saw the origin of the non-conservation of $\cE$: it can be traced back to the work done by the pressure, see Eq.~\eqref{eq:rhoPbalance}. (Even in the case $w=0$ there is a nonvanishing pressure anisotropy component $\Delta P$). The crucial difference between $\cE$ and $E m_0$ is that the former does not include the energy stored in the gravitational field.

\subsection{Non-rotating pressurised shell}

Next consider $a=0$ and $w\neq0$, in which case we must take into account the $w$-dependence of the radial potential, which is derived from Eq.~\eqref{eq:deltabeta}.
The difference between the exterior and interior gravitational masses is now given by
\be
\Delta M=M_+-M_-=\frac{m_0}{2}(\beta_+ + \beta_-)\left(\frac{m_0}{\cR^{2N}}\right)^{\frac{2N+1}{2N}w}\,,
\ee
and the total ADM mass is
\be
M_+ = M_- +m_0 \left(\frac{m_0}{\cR^{2N}}\right)^{\frac{2N+1}{2N}w} \sqrt{1+\dot{\cR}^2-\frac{2M_-}{\cR^{2N}}} - \frac{m_0^2}{2\cR^{2N}} \left(\frac{m_0}{\cR^{2N}}\right)^{\frac{2N+1}{N}w}\,.
\ee
For large $\cR$ (and assuming $w\geq0$) we get
\be
M_+ \simeq M_- +m_0 \left(\frac{m_0}{\cR^{2N}}\right)^{\frac{2N+1}{2N}w} \sqrt{1+\dot{\cR}^2}\,.
\label{eq:MlargeR}
\ee
Clearly, we must have $M_+\geq M_-$ as long as the constant $m_0^{1+\frac{2N+1}{N}w}$ is non-negative. (A scenario with $M_+<M_-$ would necessarily require $m_0^{1+\frac{2N+1}{N}w}<0$ but this violates energy conditions.)

In this case we compute the shell's energy to be
\be
\cE=\frac{\cA_{2N+1} (2N+1) m_0}{8\pi \kappa_N} \left(\frac{m_0}{\cR^{2N}}\right)^{\frac{2N+1}{2N}w}
=m_0 \left(\frac{m_0}{\cR^{2N}}\right)^{\frac{2N+1}{2N}w}\,.
\ee
We see that, if $w>0$, the intrinsic energy of the shell vanishes as it approaches infinity. This might seem suspicious at first sight, but it is in accordance with our earlier comments: the shell looses intrinsic energy as $\cR$ increases, and this goes into work done by the pressure.

\subsection{Rotating pressurised shell}

Finally, we arrive at the most general case, $a\neq0$ and $w\neq0$.
Using Eq.~\eqref{eq:deltabeta}, the difference between the exterior and interior gravitational masses is now given by
\be
\Delta M=M_+-M_-=\frac{m_0}{2}(\beta_+ + \beta_-)\left(\frac{m_0}{\cR^{2N}}\right)^{\frac{2N+1}{2N}w} \left(1+\frac{2Ma^2}{\cR^{2N+2}}\right)^{1-w/2}\,.
\ee
Once again, from this we can obtain the total ADM mass,
\beq
M_+ &=& M_- +m_0 \left(\frac{m_0}{\cR^{2N}}\right)^{\frac{2N+1}{2N}w} \left(1+\frac{2Ma^2}{\cR^{2N+2}}\right)^{\frac{1-w}{2}} \sqrt{1+\dot{\cR}^2-\frac{2M_-}{\cR^{2N}} + \frac{2Ma^2}{\cR^{2N+2}}} \nonumber\\
 && - \frac{m_0^2}{2\cR^{2N}} \left(\frac{m_0}{\cR^{2N}}\right)^{\frac{2N+1}{N}w} \left(1+\frac{2Ma^2}{\cR^{2N+2}}\right)^{1-w}\,.
\eeq
For large $\cR$ the rotation terms are subdominant and we recover Eq.~\eqref{eq:MlargeR}. 

The radial effective potential was already presented in~\eqref{eq:potential} for the general case including both rotation and isotropic pressure.
A careful inspection of its expression reveals that for $-\frac{2N}{2N+1}<w<0$ the shell is not classically allowed to be at infinity.
If $w>0$ there is no such obstacle, but if we want $\Delta M \neq0$ then the shell's velocity approaches that of light at infinity: $\dot{\cR}\sim \cR^{(2N+1)w}\to\infty$, and as we saw previously ---converting to Boyer-Lindquist coordinates--- this corresponds to $d\cR/d\cT=1$.
If $w<-\frac{2N}{2N+1}$, the shell can also be sent in from infinity, and in this case it approaches the speed of light at a rate given by
\be
\dot{\cR}\sim \frac{1}{2}\left(\cR/m_0^{1/(2N)}\right)^{-2N-(2N+1)w}\to\infty\,.
\ee

Such choices of negative $w$ can result in either plunges or bounces, and energy conditions can be satisfied or violated depending on the parameters. For full plunges in $D=5$ dimensions, the DEC is always violated for sufficiently small values of $\cR$, as in the $w\geq0$ cases. In the following section, where we present the results of the parameter scan, we limit ourselves to $w\geq0$. However, we note that in higher dimensions, $N\geq2$, there exist full plunges with $w<0$ that satisfy the dominant energy condition during the whole trajectory of the shell. This is discussed in Appendix~\ref{sec:boundsEC}.

\mysection{Scanning the parameter space
\label{sec:scan}}

In this section we will present our results concerning the outcome of the numerical scan of the parameter space describing collapses of rotating thin shells, with all independent angular momenta set equal. 

The first point to notice is that the dimensionality of the parameter space is quite large: we have a total of four continuous parameters to vary, $\{m_0, Ma^2, E, w\}$, plus one discontinuous parameter, namely the spacetime dimensionality $D=2N+3$. Even though we analyze different $N$ values independently, the dimensionality of the parameter space is too large to numerically explore entirely and, for that matter, to efficiently represent in a single figure. In order to display our results we will therefore present selected sections, by fixing values of $D$, $w$ and $E$. These sections are representative of the overall picture, and other sections scanned (but not shown) produced consistent results.
For polytropic matter shells the parameter space would be even larger.

Our analysis is concerned only with orbits that initiate at infinity. We are particularly interested in distinguishing when the shell trajectory corresponds to a full plunge, a two-world orbit or a true bounce. These trajectories are defined as follows\footnote{This classification is similar to what is done in Ref.~\cite{Diemer:2014lba} for the orbits of test particles in five-dimensional rotating black hole spacetimes.}:
\begin{itemize}
\item \textbf{Full plunges (FP):} The shell has no turning point and collapses onto the singularity.
\item \textbf{Two-world orbits (TWO):} The shell crosses a black hole horizon, has a turning point (therefore avoiding the singularity) and then exits a white hole horizon into a distinct universe. From the point of view of an asymptotic observer, the shell simply falls into the black hole, taking an infinity time to do so.
\item \textbf{True bounces (TB):} The shell has a turning point occurring outside any existing horizon (if there is one). This is a time-symmetric orbit and after reaching a minimum radius the shell disperses back to infinity.
\end{itemize}
In addition to this classification, we also evaluate explicitly whether the WEC and the DEC are satisfied throughout the shell's orbit. All the results presented in the following remain unaltered if we consider the NEC instead of the weak energy condition. The difference between them is just the inclusion or not of a single inequality ($WEC_0\geq0$), which does not impose further constraints on the trajectories scanned.

The results presented below were obtained with two complimentary methods. The most straightforward one ---but also the more computationally intensive--- envolves directly sweeping though the selected sections of the parameter space (by varying $m_0$ and $Ma^2$) and evaluating the quantities of interest. The other ---more efficient--- strategy is to obtain directly the curves that separate different regions in the phase space. For example, the lines marking the boundary between full plunges and bounces are computed by imposing that a local maximum of the effective potential takes the value $V_{eff}=0$. This corresponds to the critical configurations we are looking for: small changes in the parameters can raise or lower the potential barrier above or below zero, yielding a bounce or a plunge trajectory, respectively.

The fact that we obtained fully consistent results with the two approaches serves as a good check on our calculations. 

A word of caution is in order. In dimensions $D\geq7$, and for some choices of the parameters, our shell evolutions can originate highly spinning black holes (nearly extremal) that are known to be unstable~\cite{Dias:2010eu,Dias:2011jg}. In these extreme cases, the resulting cohomogeneity-1 Myers-Perry black hole cannot be expected to be the endpoint of the collapse.

\subsection{Rotating dust shells starting from rest at infinity: $w=0$, $E=1$, varying $D$}

As pointed out in~\cite{Delsate:2014iia}, having the thin shell initially at rest at infinity requires that we consider $w=0$ and $m_0=\Delta M=M_+-M_-$, at least when restricting to linear EoS. In this case the effective potential reduces to
\be
V_{\rm eff}(\cR) 
= 1 + \frac{2Ma^2}{\cR^{2N+2}} - \frac{2M_-+m_0}{\cR^{2N}}
 - \left(1+\frac{2Ma^2}{\cR^{2N+2}}\right)^{-1}
 -\frac{1}{4} \left(\frac{m_0}{\cR^{2N}}\right)^2 \left(1+\frac{2Ma^2}{\cR^{2N+2}}\right).
\label{eq:potential3}
\ee

Note that there is a scale invariance in the problem: the shell's equation of motion remains unchanged when the masses $M_-$ and $m_0$ are rescaled by a factor $\kappa$, while the radius $\cR$, the spin $a$ and the proper time $\tau$ are rescaled by a factor $\kappa^{1/(2N)}$. As a consequence we can, without loss of generality, set $M_-=1$.
This allows us to reduce the dimensionality of the parameter space to just 2: these collapses depend only on $m_0$ and $Ma^2$, up to a trivial rescaling.

\begin{figure}[!t]
\centering
\subfigure{\includegraphics[width=0.40\textwidth]{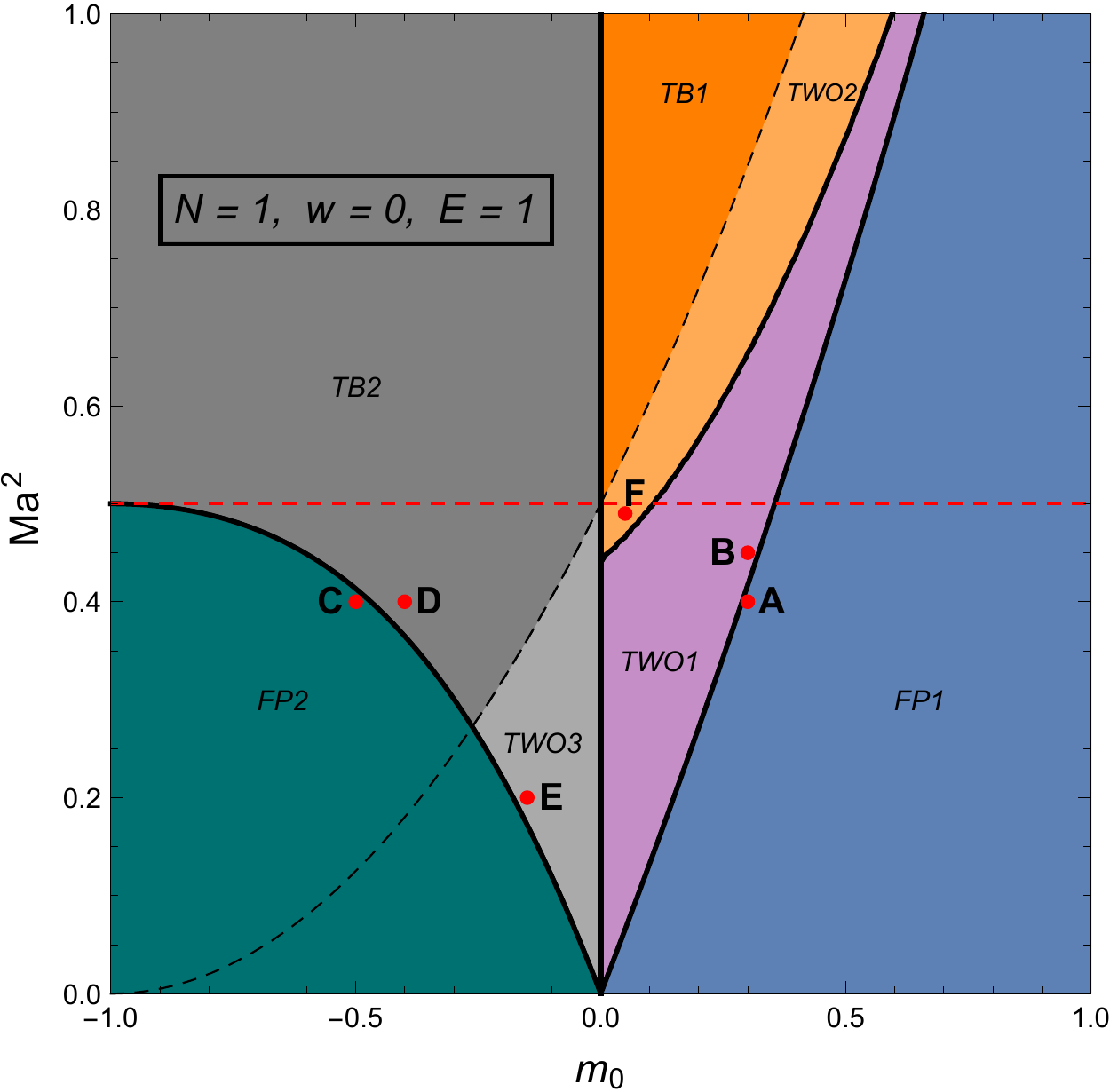}}
\quad
\subfigure{\includegraphics[width=0.40\textwidth]{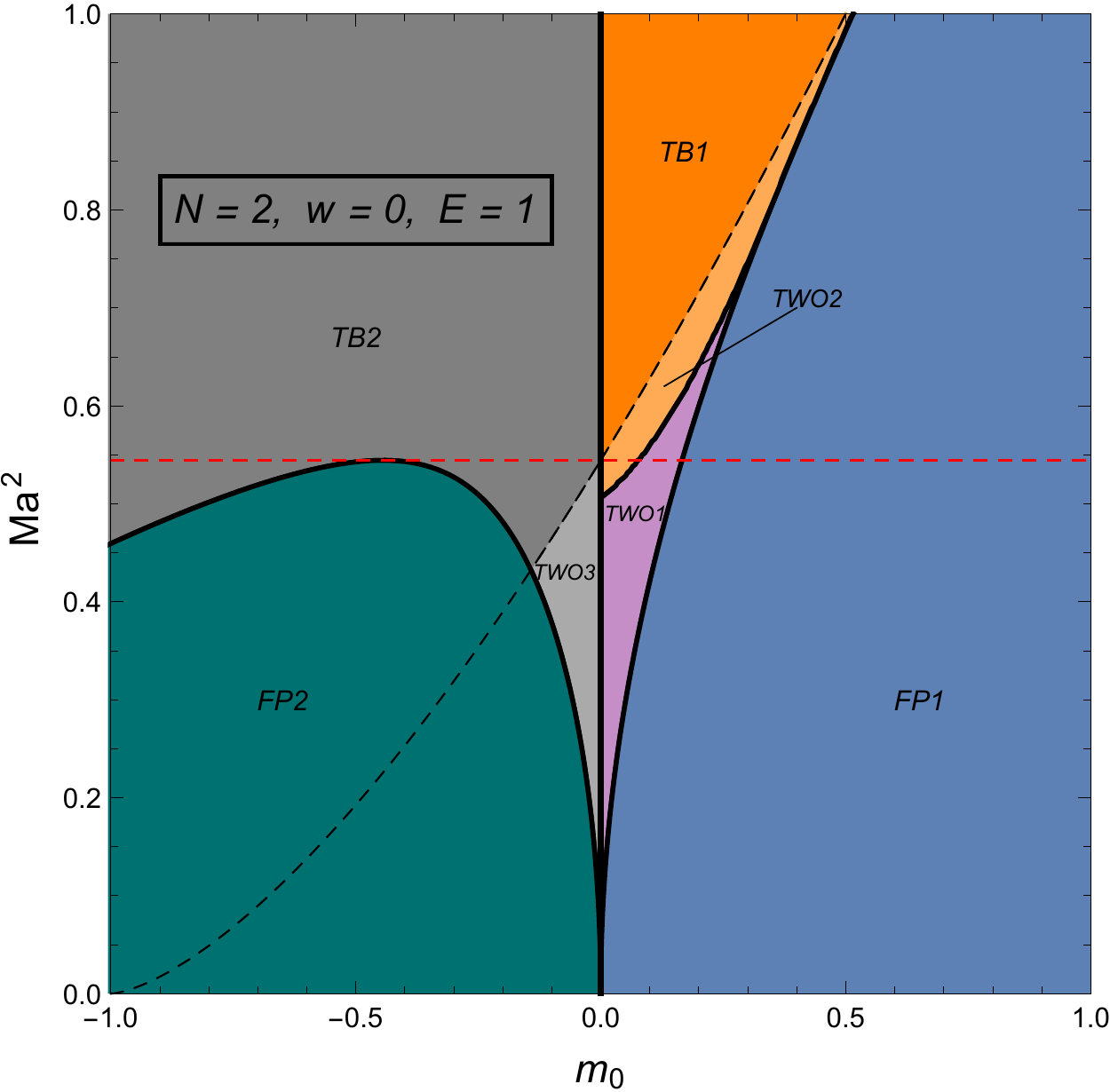}}
\\
\subfigure{\includegraphics[width=0.40\textwidth]{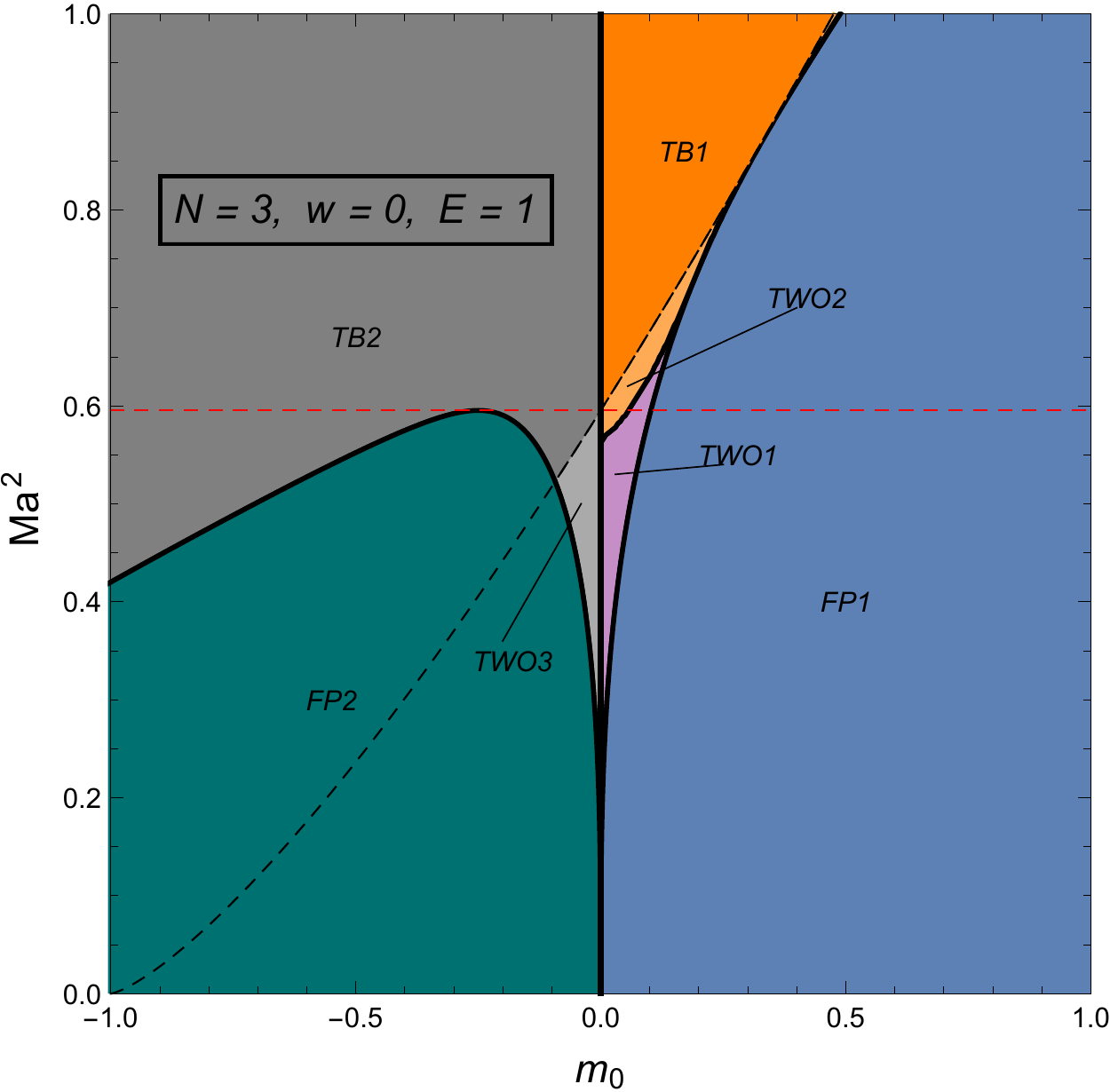}}
\quad
\subfigure{\includegraphics[width=0.40\textwidth]{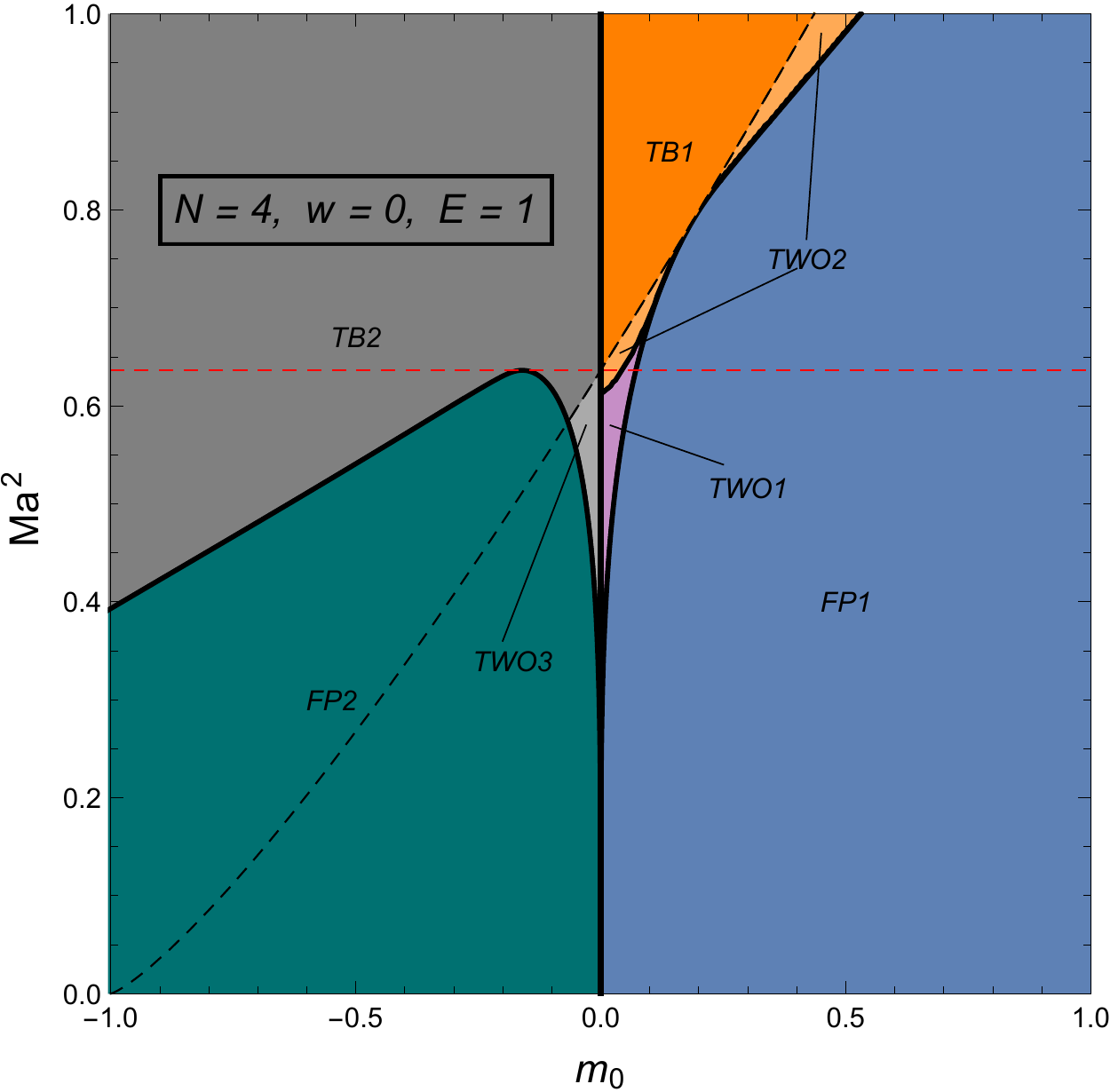}}
\caption{Plots of the $\{m_0,Ma^2\}$ parameter space fixing $E=1$ and $w=0$.  We display results for four different spacetime dimensions, namely $D=5, 7, 9, 11$, corresponding to $N=1, 2, 3, 4$. The blue regions (FP1) indicate full plunges satisfying the WEC (though all of them violate the DEC), while the green regions (FP2) correspond to full plunges violating the WEC. Light-purple regions (TWO1) indicate two-world orbits satisfying the WEC but not the DEC, and light-orange regions (TWO2) correspond to two-world orbits satisfying the WEC and the DEC. True bounces respecting both the WEC and the DEC are indicated by dark-orange domains (TB1). The light-gray regions (TWO3) correspond to two-world orbits that violate the WEC and those colored dark-gray (TB2) identify true bounces violating the WEC. The dashed black curve indicates the maximum value of $Ma^2$ for which the external geometry possesses an horizon (and therefore corresponds to an extremal black hole). The horizontal dashed red line indicates a similar situation but for the interior geometry. The points marked \textbf{A--F} in the first panel were chosen each one in a different region, and the respective plots showing the potential, as well as the WEC and DEC curves, are displayed in Fig.~\ref{fig:potentials}.}
\label{fig:w0_E1_varyN}
\end{figure}

In Appendix~\ref{sec:boundsEC} it is shown that under these conditions ($w=0$ and $m_0=\Delta M$) and imposing $Ma^2<(Ma^2)_{max}$ so that the interior spacetime has an event horizon, the WEC is satisfied (violated) if $m_0>0$ ($m_0<0$). The same thing holds for condition $DEC_1$, but $DEC_\alpha$ is violated at sufficiently small $\cR$, which is necessarily explored by full plunges ---though this occurs always inside the exterior horizon, when there exists one.

In Fig.~\ref{fig:w0_E1_varyN} we display our results for the scan of the parameter space $\{m_0,Ma^2\}$, for the four lowest spacetime dimensionalities we can consider: $D=5,7,9,11$. Different regions are identified according to the classification above (full plunges, two-world orbits or true bounces), indicating also whether the geometries interior and exterior to the shell correspond to a black hole (below the red and black dashed curves) or a naked singularity (above the red and black dashed curves).

We allow the shell's proper mass to be negative but, as indicated before, regions with $m_0<0$ violate the WEC for all values of $Ma^2$, while for $m_0>0$ the WEC is always satisfied, so the vertical black line at $m_0=0$ divides the plots in two regions, and the right-hand half of the plots corresponds to physically reasonable matter content on the shell.

Fig.~\ref{fig:w0_E1_varyN} also provides visual confirmation of a point made above: the blue (FP1) and the orange regions (TB1 and TWO2) never intersect, i.e., for these cases ($w=0$) full plunges always violate the DEC, even when the WEC is satisfied.
This is more obvious in the first panel, $D=5$, where domains FP1 and TWO2 are separated by an intermediate light-purple region, TWO1. For higher dimensions, $N\geq2$, it is clear that this region extends only up to a finite value of $m_0$, beyond which the full plunge region FP1 and the two-world orbit region TWO2 become contiguous.

There are two notable conclusions that can be inferred from Fig.~\ref{fig:w0_E1_varyN}, especially from the last three panels. One is that true bounces only occur when the geometry exterior to the shell corresponds to a naked singularity, i.e., above the dashed black line. This will change when we consider shells with $w\neq0$. Another is that full plunges satisfying the WEC always form an exterior horizon, while full plunges violating the WEC ($m_0<0$) only occur when the interior geometry has a black hole. The marginal orbits lying at the border of the full plunge regions FP1 and FP2 become tangent at one point to the black and red dashed curves, respectively. These special points indicate the appearance of a turning point that coincides with the black hole horizon radius.

In all panels, the physically more interesting quadrant is the lower-right one, where the WEC is satisfied and the geometry interior to the shell is dressed. For the case $w=0$ we are considering, no true bounces can be found in this quadrant. Note that as the dimensionality $N$ increases, more of this region is covered by full plunges; for $N=4$ only a small corner corresponds to TWOs, and the tiny region where the DEC is satisfied occurs only for shells falling through near extremal black holes.

\begin{figure}[p]
\centering
\subfigure{\includegraphics[width=0.38\textwidth]{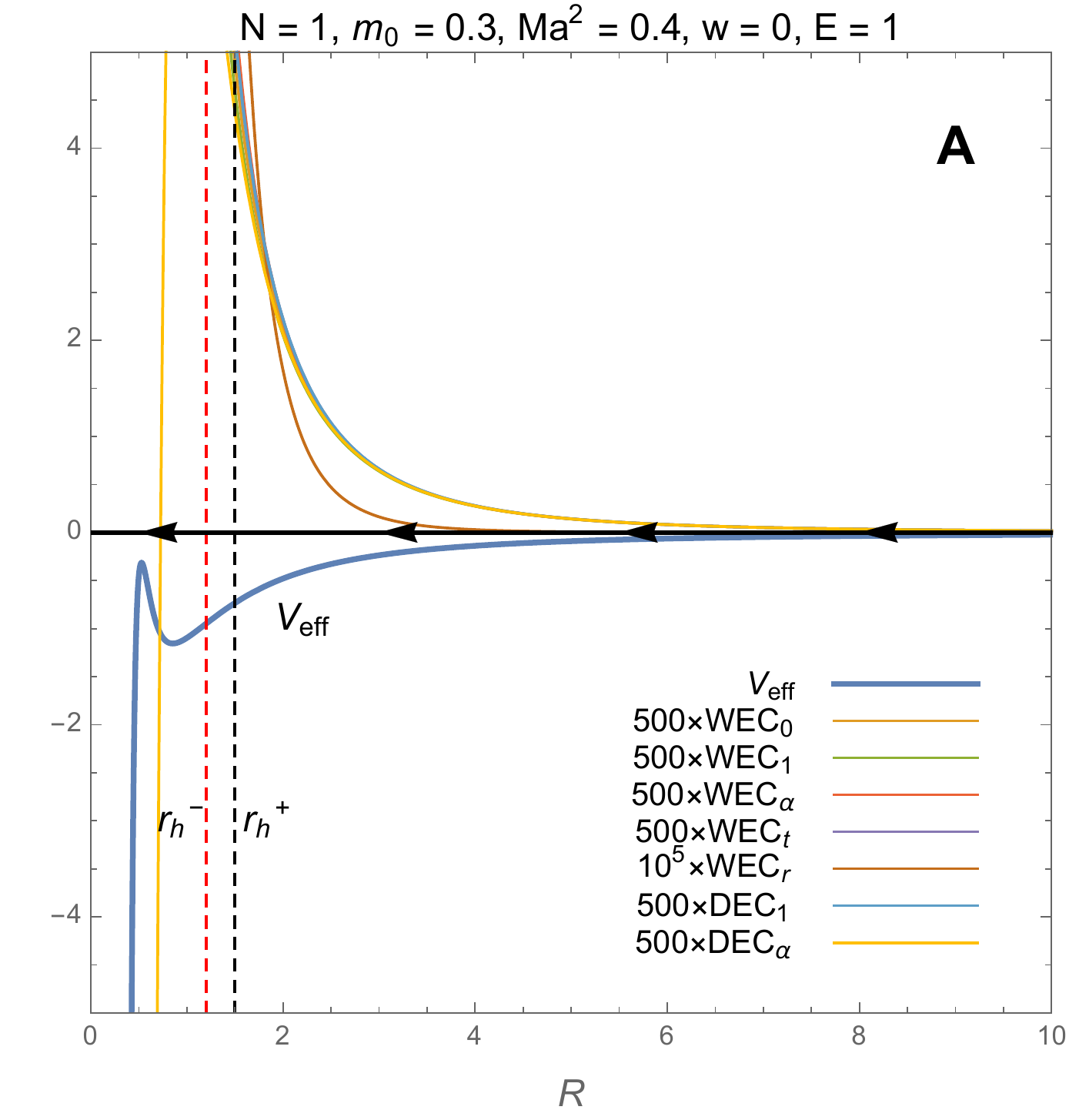}}
\quad
\subfigure{\includegraphics[width=0.38\textwidth]{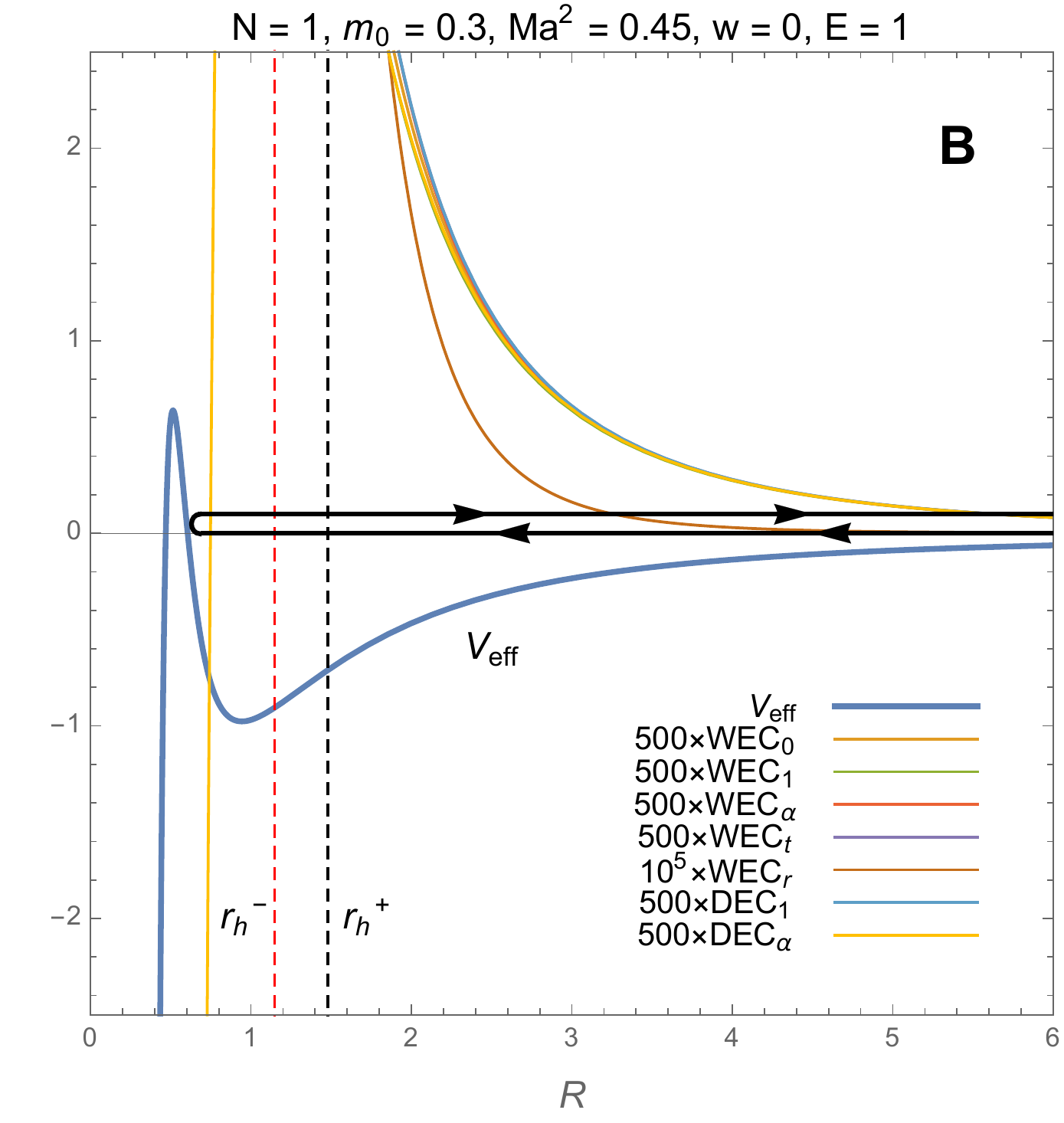}}
\\
\subfigure{\includegraphics[width=0.38\textwidth]{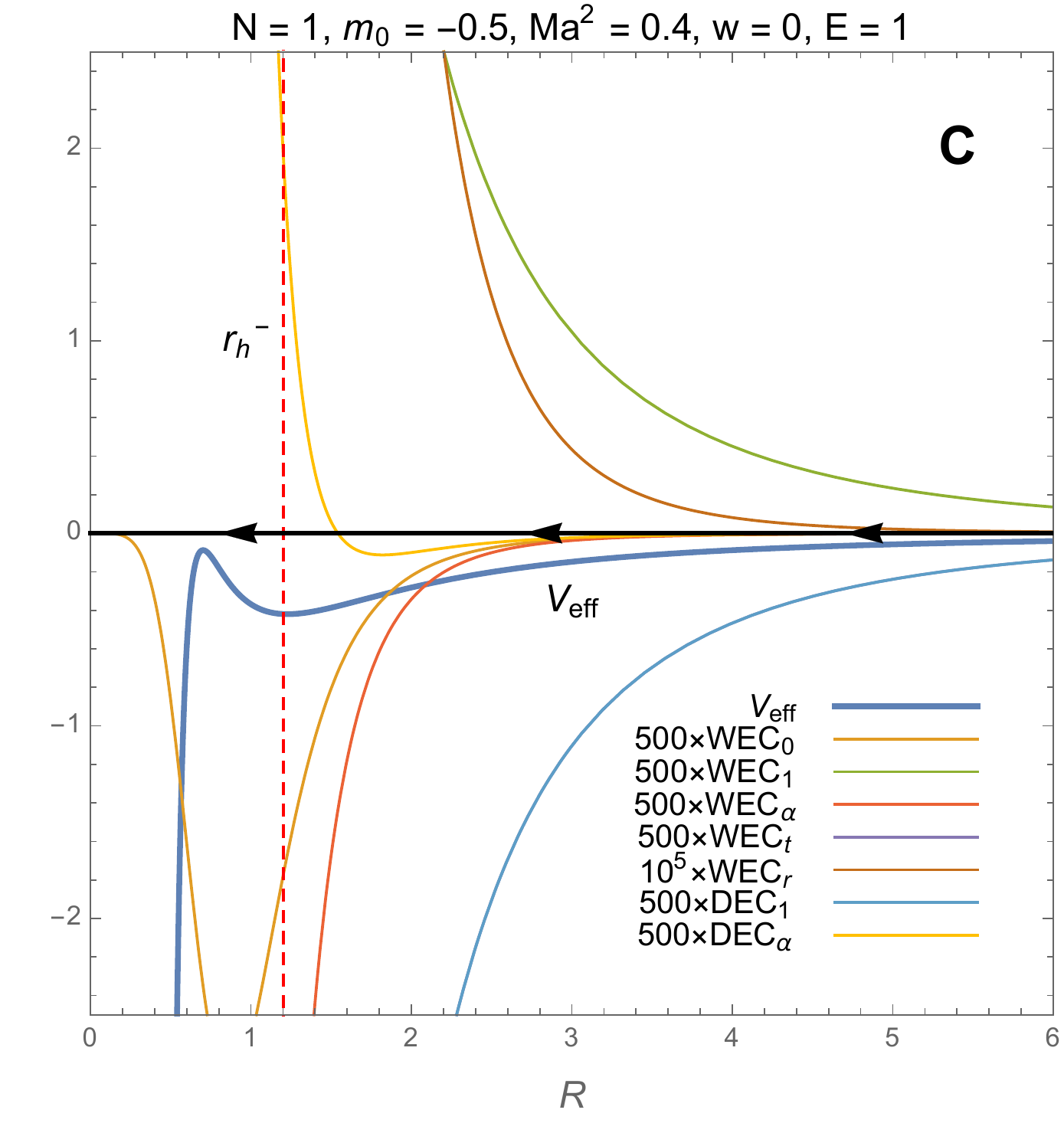}}
\quad
\subfigure{\includegraphics[width=0.38\textwidth]{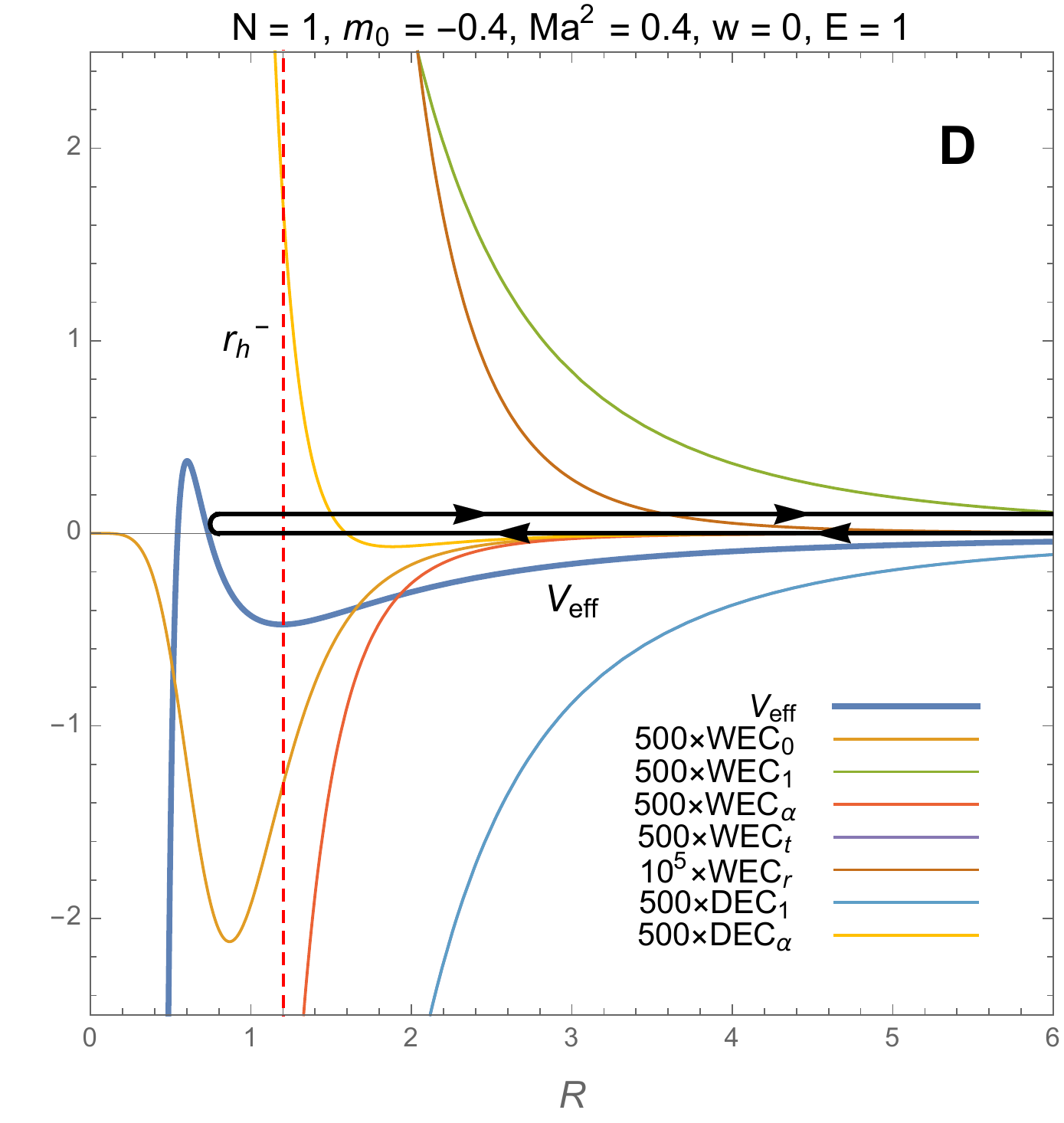}}
\\
\subfigure{\includegraphics[width=0.38\textwidth]{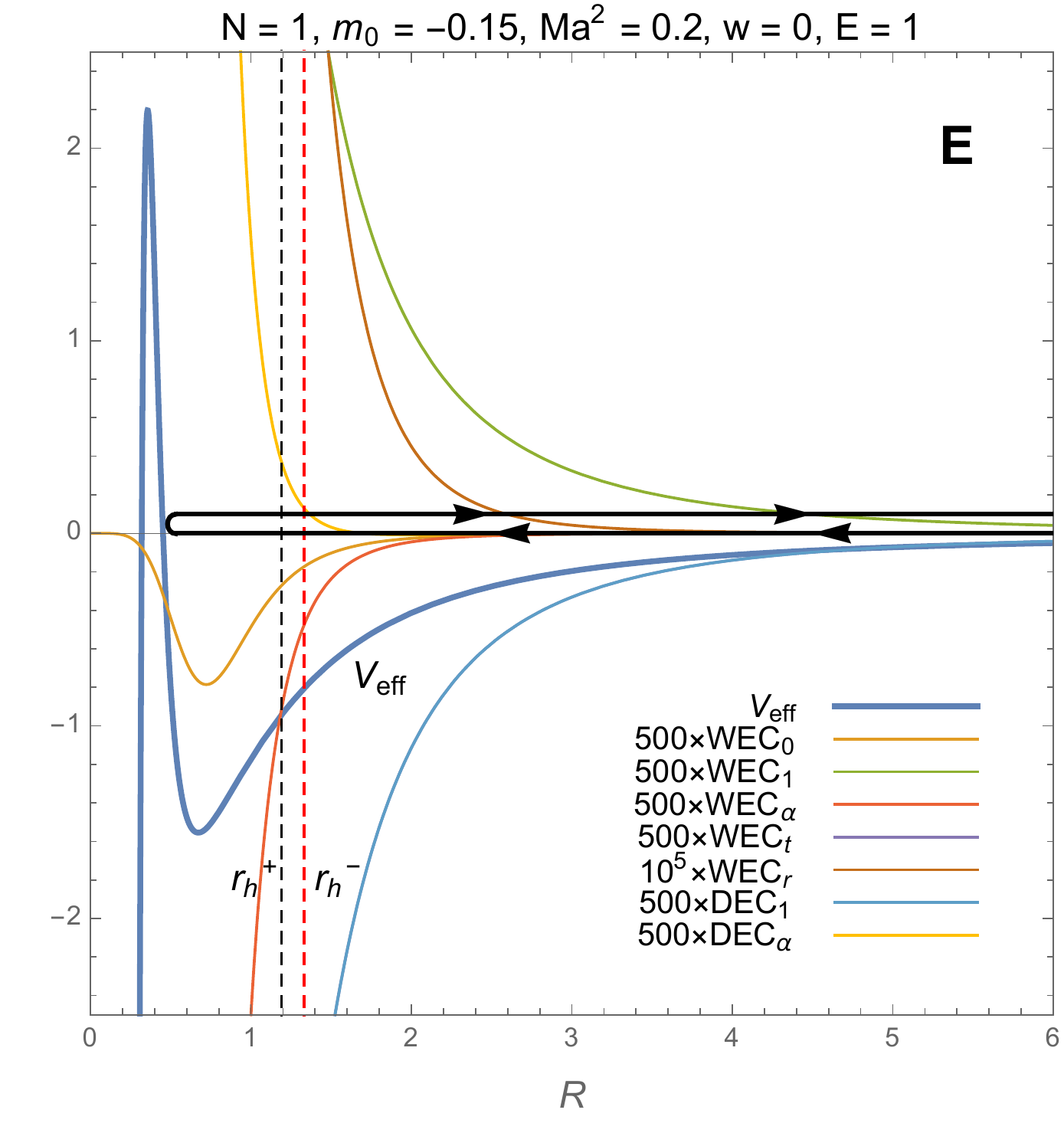}}
\quad
\subfigure{\includegraphics[width=0.38\textwidth]{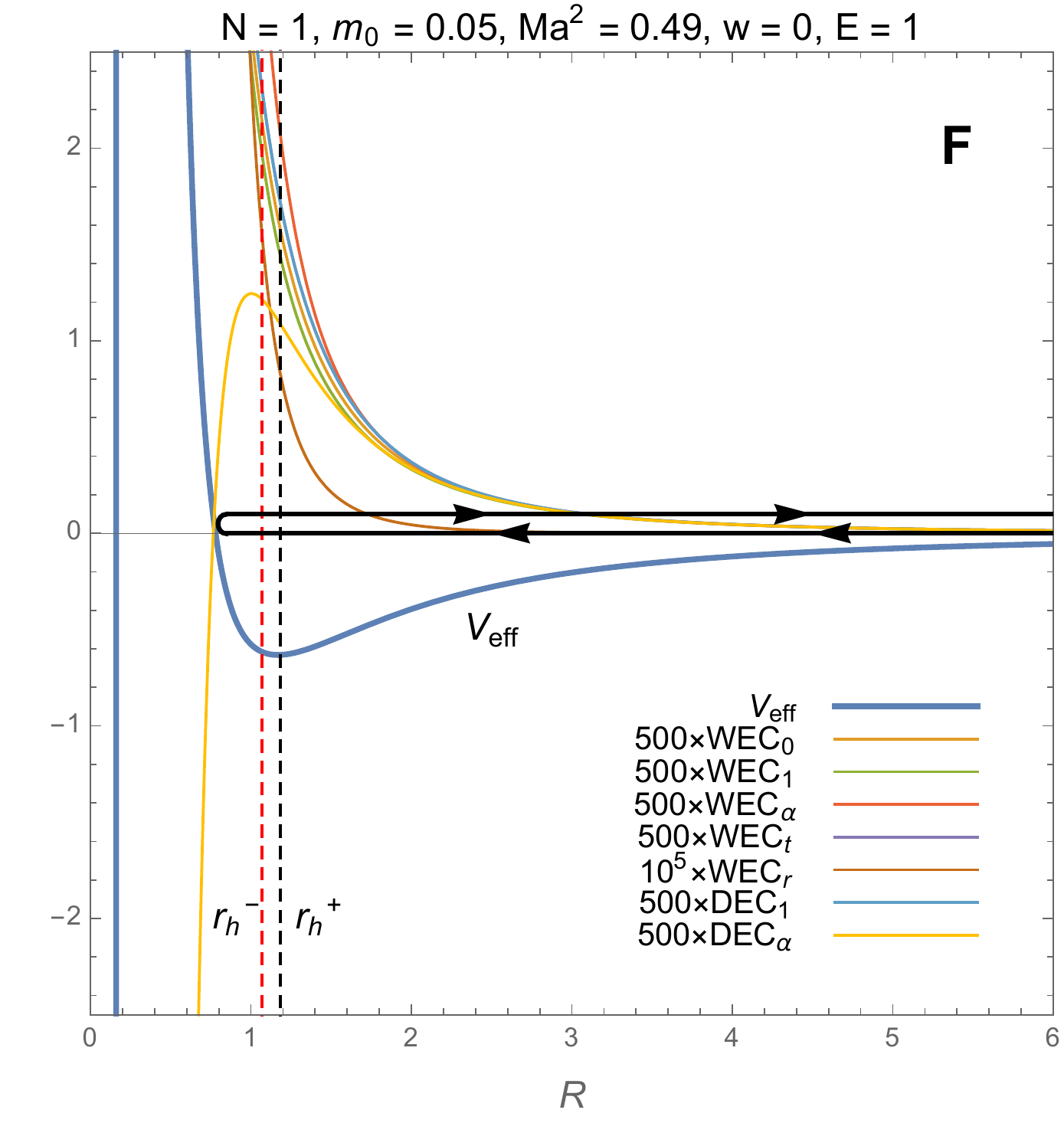}}
\caption{Plots of the radial potential $V_{eff}$ and the various energy conditions~\eqref{eq:WEC} and~\eqref{eq:DEC}, corresponding to choices of parameters indicated by the points \textbf{A--F} in the first panel of Fig.~\ref{fig:w0_E1_varyN}. The various WEC and DEC conditions are satisfied only when the respective graphs are positive. Panels \textbf{A} and \textbf{B} represent a full plunge and a two-world orbit, respectively, both with WEC satisfied and $DEC_{\alpha}$ violated inside the horizons. Panels \textbf{C} and \textbf{D} correspond to a full plunge and a true bounce (without horizon in the exterior), respectively, both with WEC violated. Panel \textbf{E} represents a two-world orbit with WEC violated, and panel \textbf{F} shows a two-world orbit in which both WEC and DEC are satisfied (note that $DEC_{\alpha}$ would be violated only inside the turning point). The vertical dashed lines indicate the locations of the horizons for the geometries exterior (black) and interior (red) to the shell.}
\label{fig:potentials}
\end{figure}

At last, we comment on the implications of our results regarding cosmic censorship. From this point of view, the potentially dangerous situation is the one in which the shell fully collapses onto a pre-existing black hole (below the red dashed line) and ends up with a naked singularity (above the black dashed line). This corresponds to the FP2 region, which necessarily violates the WEC. The domain TB2 also starts off with a black hole in the interior and an over-extreme exterior, and it describes the temporary appearance of a naked singularity, followed by the re-creation of the horizon after the shell bounces. In any case, the WEC is also violated for TB2, since the two dashed lines cross exactly at $m_0=0$.
In summary, these results are in accordance with the weak cosmic censorship conjecture.

\subsection{Rotating dust shells with radial velocity at infinity: $D=5$, $w=0$, varying $E$}

Having studied the effect of dimensionality on the space of collapses, we will now fix $N=1$ for the remainder of the paper, i.e., we consider a five-dimensional spacetime.
In this subsection we allow for the infalling shell to start with {\em finite} velocity at infinity, by varying the energy parameter $E>1$.
For concreteness, we take $w=0$, corresponding to shells with vanishing isotropic pressure component.

\begin{figure}[!t]
\centering
\subfigure{\includegraphics[width=0.40\textwidth]{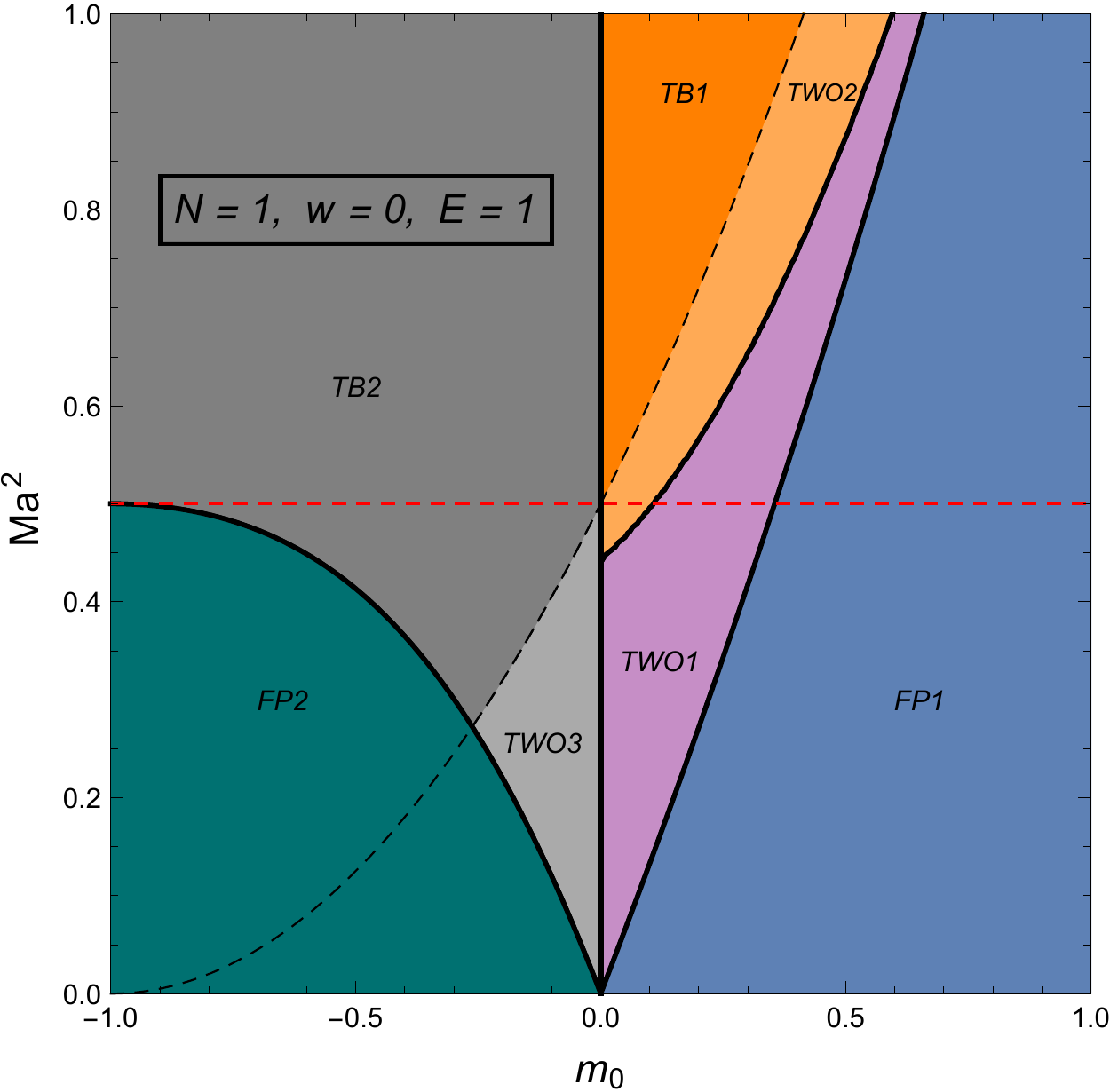}}
\quad
\subfigure{\includegraphics[width=0.40\textwidth]{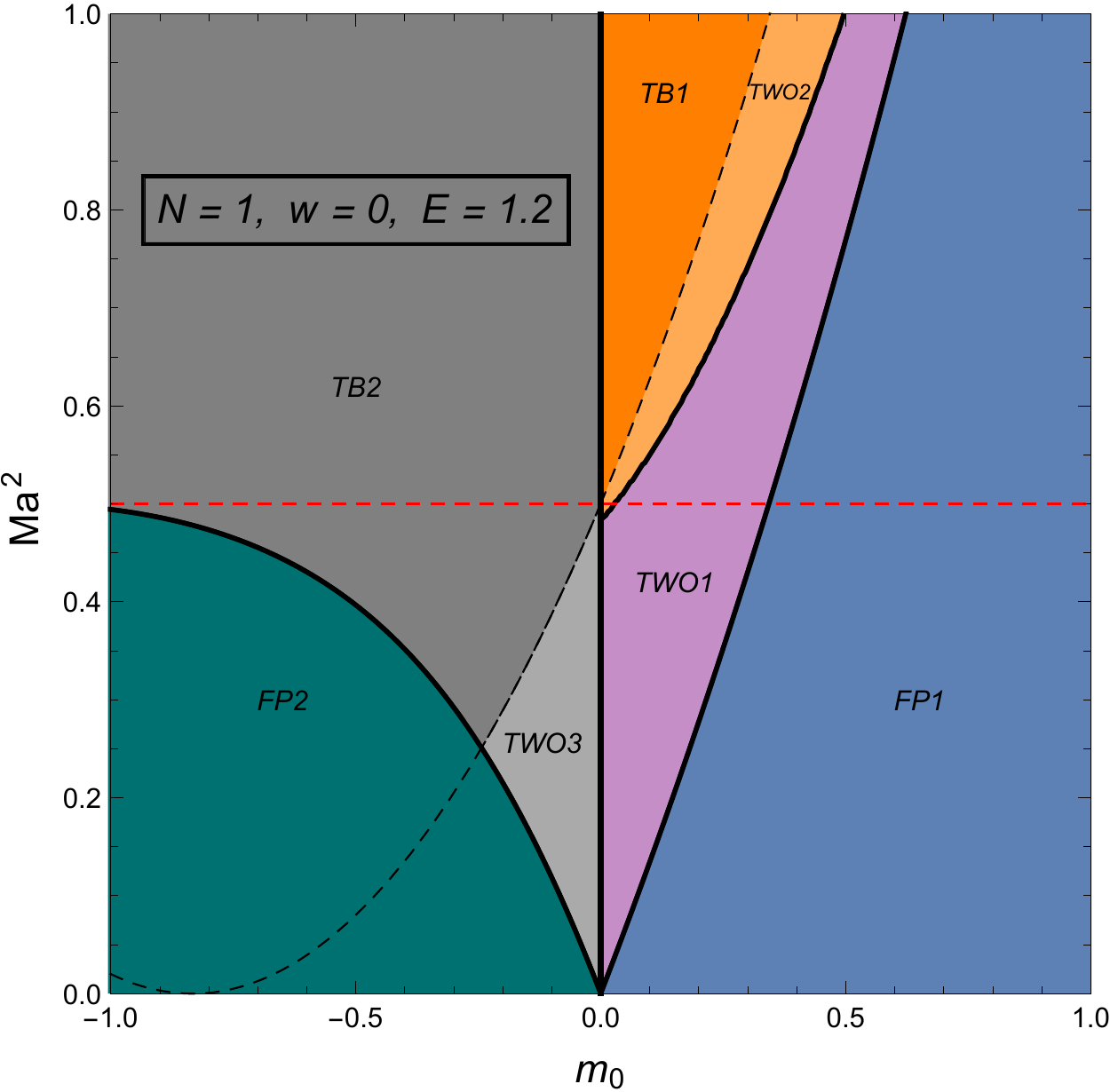}}
\\
\subfigure{\includegraphics[width=0.40\textwidth]{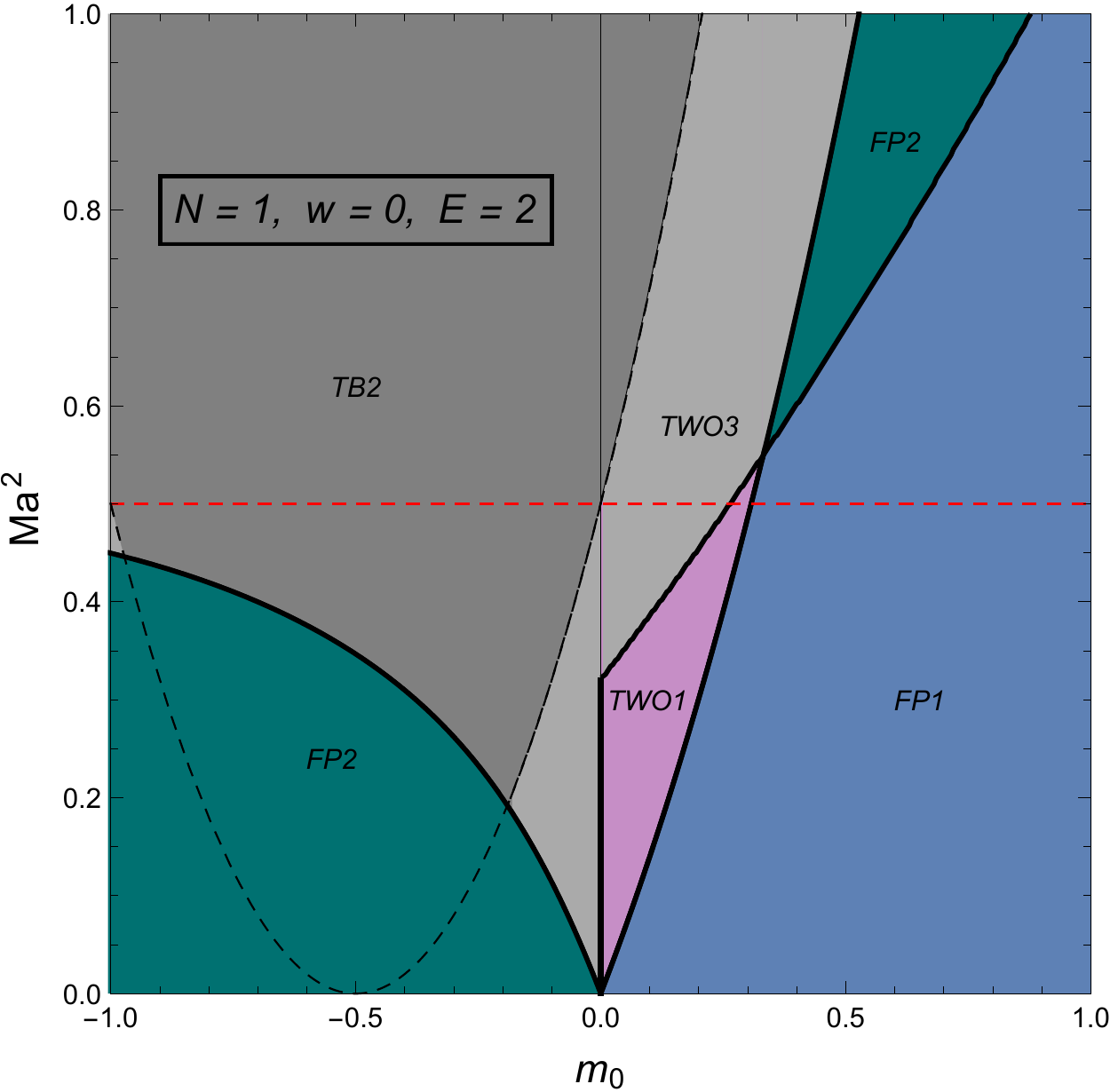}}
\quad
\subfigure{\includegraphics[width=0.40\textwidth]{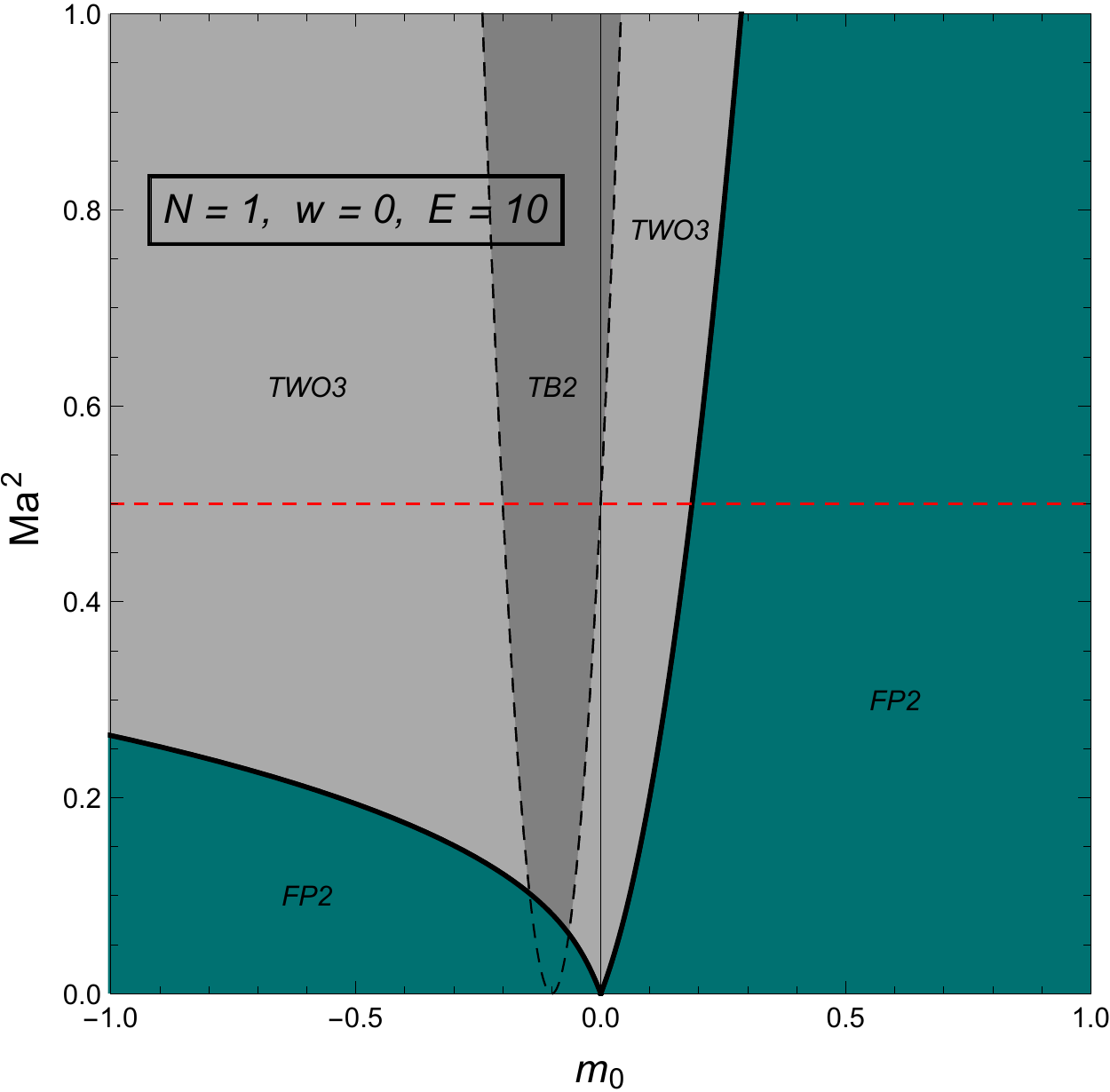}}
\caption{Plots of the $\{m_0,Ma^2\}$ parameter space fixing $N=1$ ---corresponding to five-dimensional spacetimes--- and $w=0$.  We display results for four different choices of the energy parameter $E=1, 1.2, 2, 10$. The top left panel is a repetition of the first panel of Fig.~\ref{fig:w0_E1_varyN} but we include it here to ease the comparison. The remaining three values translate into the shell having finite velocity at infinity, given respectively by $55.3\%, 86.6\%$ and $99.5\%$ the speed of light. The color coding and region names are the same as in Fig.~\ref{fig:w0_E1_varyN}.}
\label{fig:N1_w0_varyE}
\end{figure}

For our calculations it is convenient to work with the energy parameter $E$, since it directly connects the ADM mass of the exterior spacetime $M_+$ with the proper mass of the shell $m_0$ [see Eq.~\eqref{eq:MEm0}]. However, for presentation purposes it is useful to translate this into a more intuitive quantity, such as the velocity of the shell at infinity, expressed as a fraction of the speed of light. The relation between the two is straightforwardly obtained using~\eqref{eq:VhatVeff} and~\eqref{eq:EdRdtau},
\be
\left|\frac{d\cR}{d\cT}\right|_{\cR\to\infty} = \frac{\sqrt{E^2-1}}{E}\,.
\ee
Therefore, the previous case $E=1$ indeed corresponds to shells starting from rest at infinity, but $E=1.2$ already yields a radial velocity at infinity as large as $55.3\%$ the speed of light.

The first significant difference between this case and the previous subsection is that now the WEC can be violated even for $m_0>0$, as can be seen in the last two panels of Fig.~\ref{fig:N1_w0_varyE}. For $E=10$ all the shell trajectories in the parameter space scanned violate the WEC.

As the energy parameter $E$ is increased, the region corresponding to plunges with positive proper mass $m_0$ grows, as expected. Raising the value of $E$ also shrinks the region where the DEC is satisfied. For $E=2,10$ it is not satisfied anywhere in the scanned space. Notice that the minimum of the black dashed parabola moves to the right as $E$ increases, while it always intersects the red dashed horizontal line ($Ma^2=1/2$) at $m_0=0$. As a result, the region corresponding to true bounces shrinks, and for very large values of $E$ it gets squeezed into a small interval around $m_0=0$.

Recall the region of interest for the cosmic censorship conjecture is below the red dashed line and above the black dashed line, which gets smaller as $E$ is increased. Once again, collapsing shells onto black holes result in naked singularities only in regions where the WEC is violated, in accordance with cosmic censorship.

\subsection{Rotating pressurised shells ($D=5$, varying $w$ and $E$)}

For the more general case of matter shells with a nonvanishing isotropic pressure component, $w\neq0$, it is not possible to have the shell starting from rest at infinity. Also there is no reason to impose $m_0=\Delta M$, so there is one more free parameter, namely $E$. So now we have a four-dimensional parameter space: $\{m_0, Ma^2, w, E\}$.

It can be shown that the weak energy condition is violated for EoS parameter $w>(2N+1)^{-1}$. This is due to some eigenvalues of the stress-energy tensor becoming imaginary for sufficiently large $\cR$, see Appendix~\ref{sec:boundsEC}. It can also be shown that $DEC_\alpha \geq0$ is violated at sufficiently small $\cR$ unless $w<-1/N$.
In what follows we focus our attention on positive $w$, but in Appendix~\ref{sec:boundsEC} we briefly consider the case of negative $w$, which translates into a negative pressure $P$, i.e., a {\em tension}. This case is interesting because the tension of the shell will assist the gravitational collapse, creating what might seem to be more favorable conditions to destroy the horizon and form a naked singularity. Nevertheless, there is only a narrow window allowing for negative $w$ shells to be thrown from infinity, while satisfying the dominant energy condition,
\be
-1\leq w < \min\left\{-\frac{2N}{2N+1},-\frac{1}{N}\right\}\,.
\label{eq:wnarrow}
\ee
The lower bound is derived from $WEC_t\geq0$, see Eq.~\eqref{eq:WEC}. The upper bound comes from either $V_{eff}(\cR\to\infty)\leq0$ or $DEC_\alpha|_{\cR\to0}\geq0$, see Appendix~\ref{sec:boundsEC}.
In any case, we never observe naked singularity formation from the collapse of shells (satisfying the DEC) onto black holes.

\begin{figure}[p]
\centering
\subfigure{\includegraphics[width=0.40\textwidth]{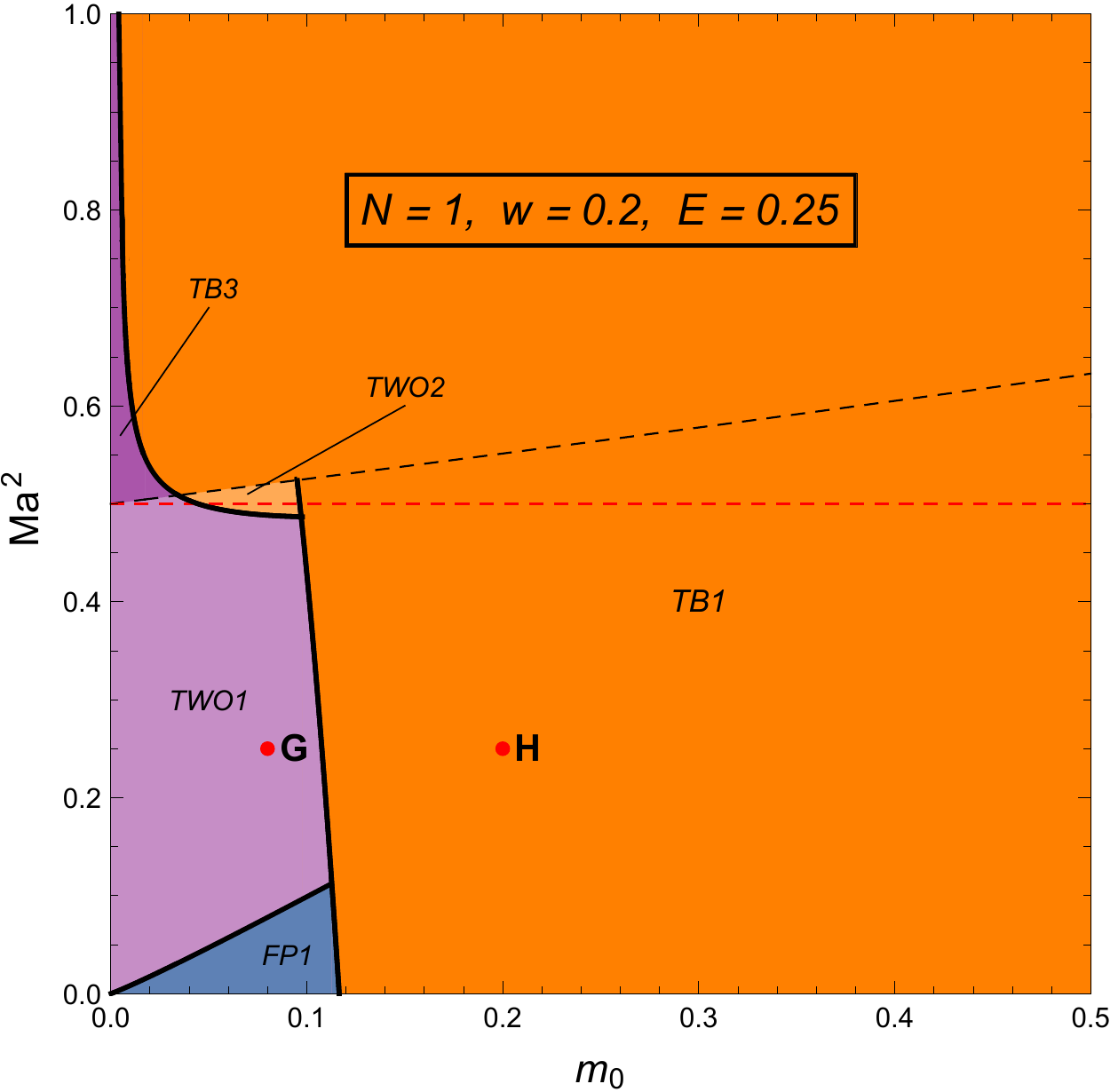}}
\quad
\subfigure{\includegraphics[width=0.40\textwidth]{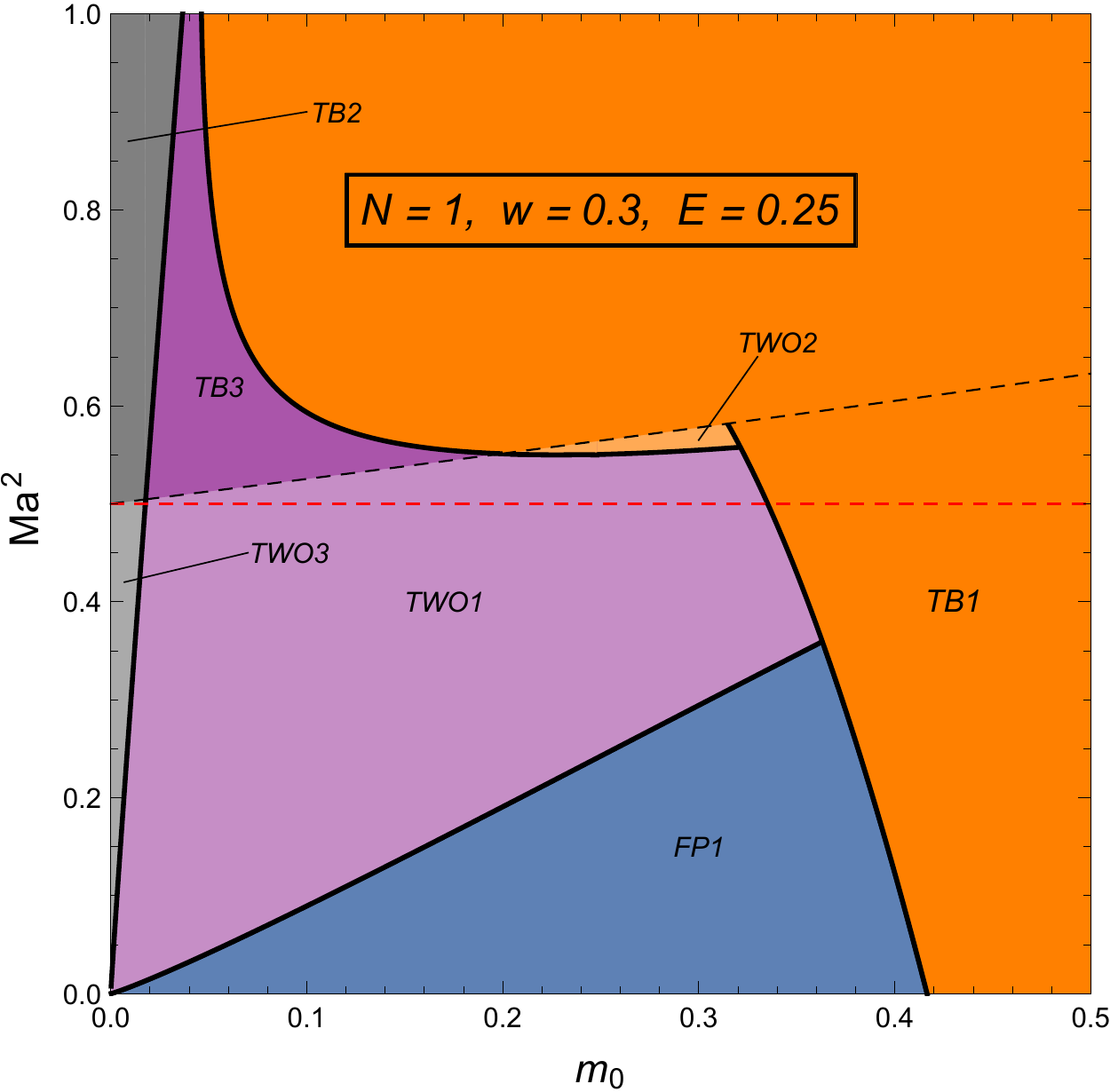}}
\\
\subfigure{\includegraphics[width=0.40\textwidth]{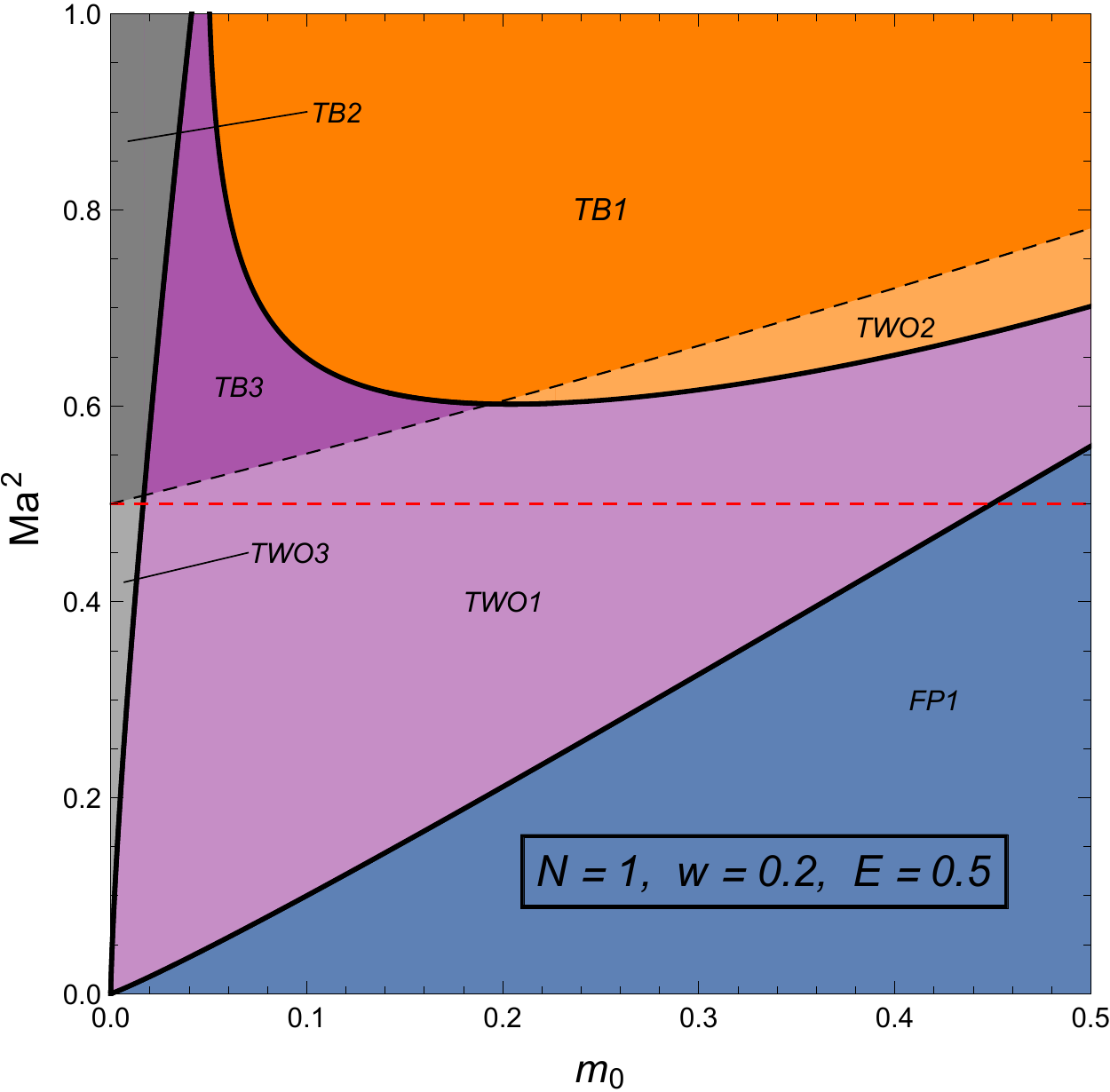}}
\quad
\subfigure{\includegraphics[width=0.40\textwidth]{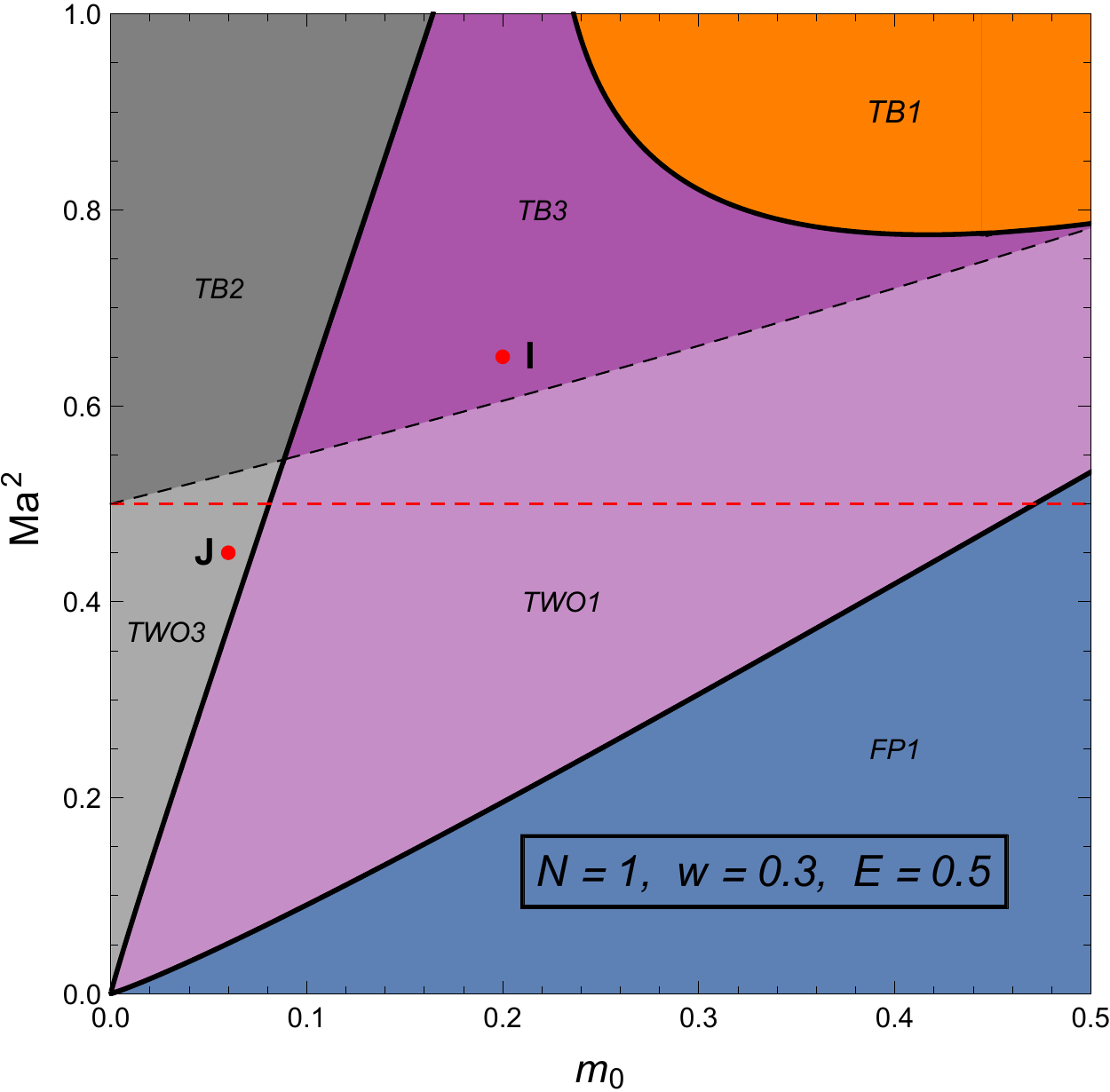}}
\\
\subfigure{\includegraphics[width=0.40\textwidth]{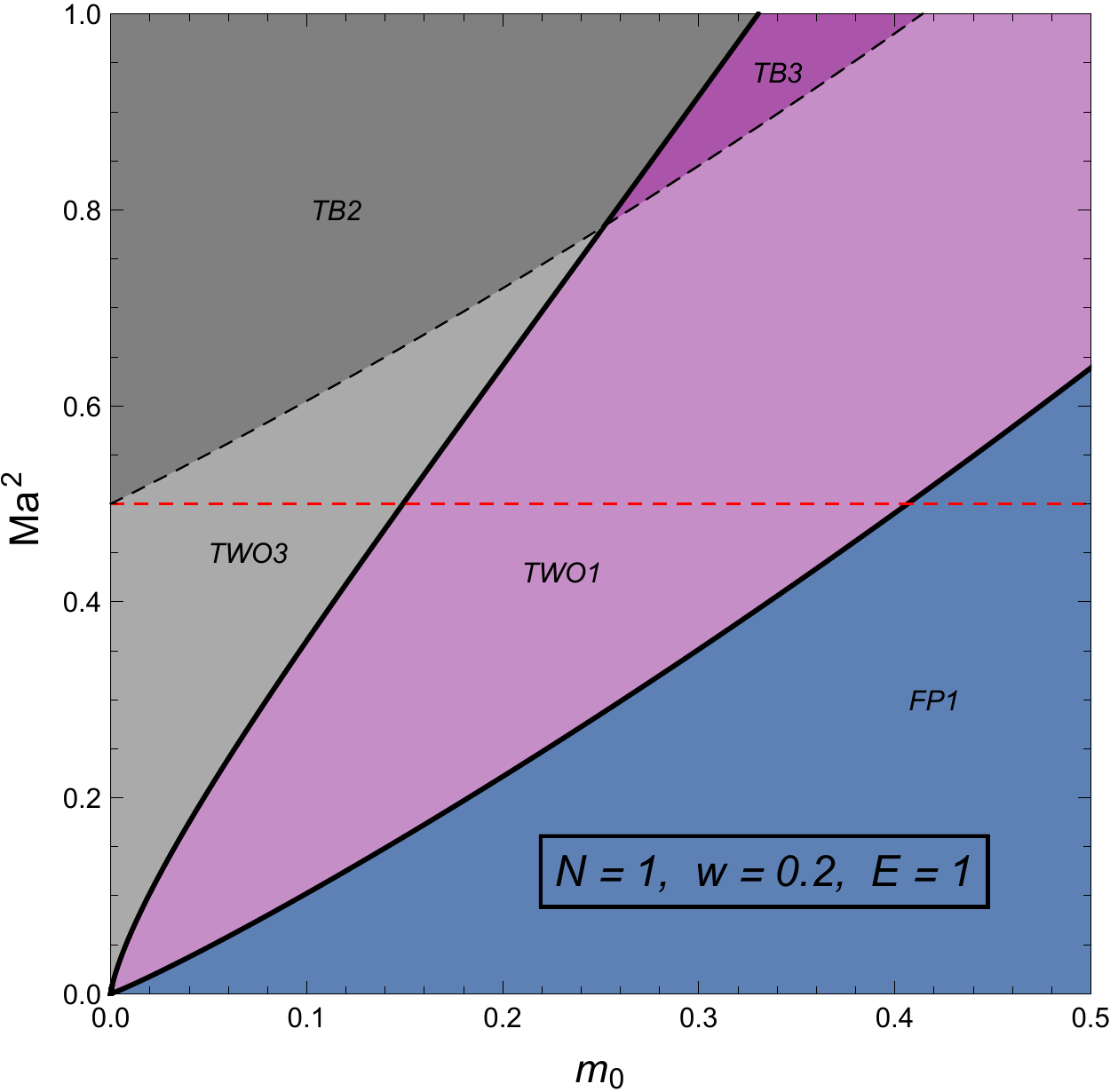}}
\quad
\subfigure{\includegraphics[width=0.40\textwidth]{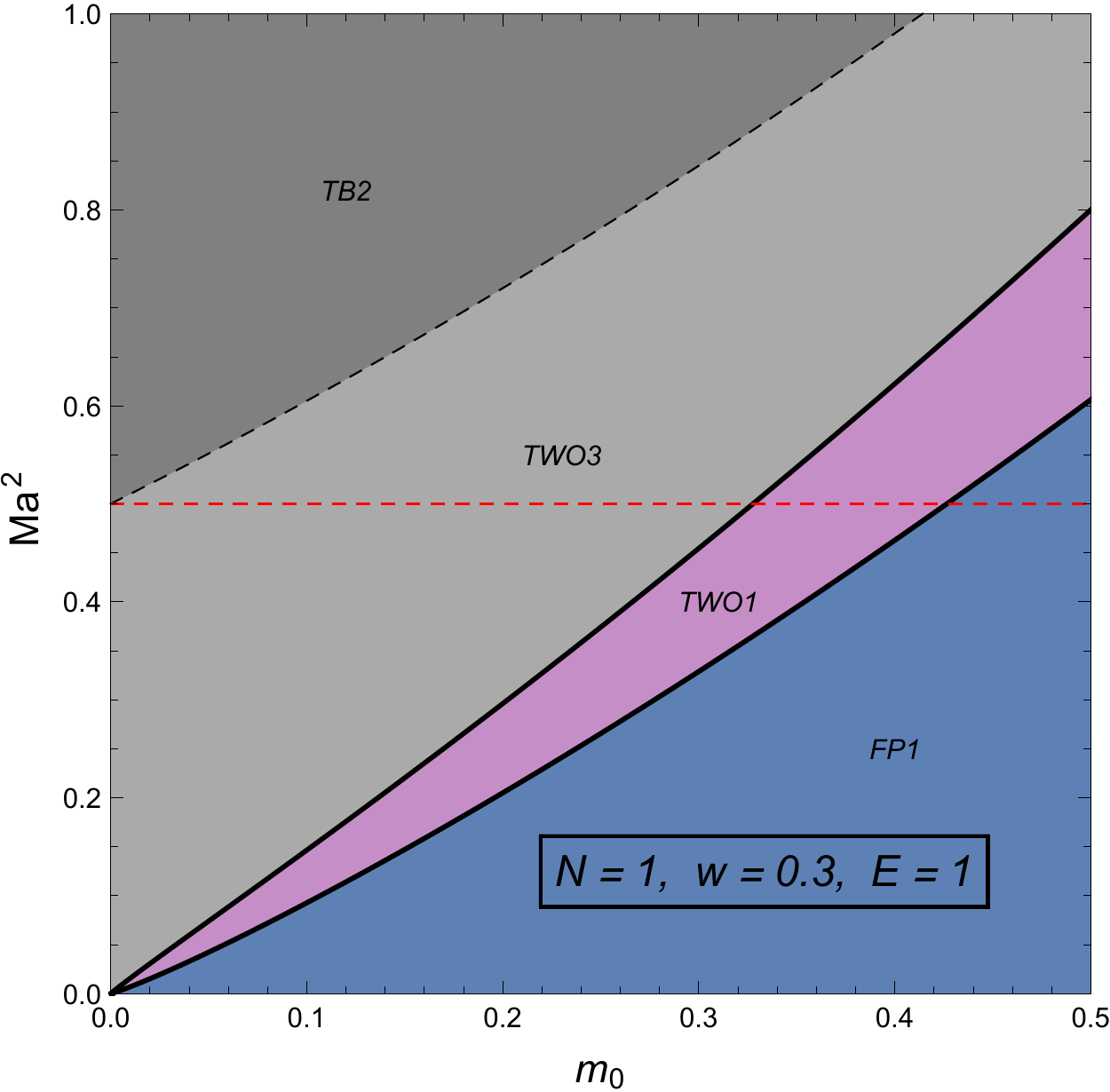}}
\caption{Plots of the $\{m_0,Ma^2\}$ parameter space fixing $N=1$, corresponding to five-dimensional spacetimes.  All combinations with $w=0.2, 0.3$ and $E=0.25, 0.5, 1$ are shown. The color coding and region names are the same as in Fig.~\ref{fig:w0_E1_varyN}, with the addition of a new dark-purple region (TB3) that indicates a true bounce satisfying the WEC but not the DEC.}
\label{fig:N1_varyw_varyE}
\end{figure}

It is evident, from a glance at Eqs.~\eqref{eq:deltabeta} and~\eqref{eq:potential}, that a nonzero $w$ easily leads to non-integer exponents of $m_0$ in the radial potential and in the energy conditions. For this reason, here we restrict the scanned region to $m_0 >0$. Otherwise, not even the weak energy condition could be satisfied with the shell at infinity.  Just like in the cases $w=0, E>1$ studied above, we find again an area with $m_0 >0$ for which the WEC is violated, as can be seen in the Fig.~\ref{fig:N1_varyw_varyE}. We only present results for $w=0.2$ and $w=0.3$ because the case $w=0.1$, with the choice of energy parameter $E=0.25$, results in an uninteresting parameter space almost entirely filled by the TB1 region, with very small domains corresponding to full plunges and two-world orbits. Nevertheless, the plots presented are sufficient to infer the trend followed when varying the EoS parameter $w$. The points \textbf{G--J} marked in some panels of Fig.~\ref{fig:N1_varyw_varyE} were chosen as representatives of four different regions, and the respective plots of the potential, as well as the WEC and DEC constraints, are shown below in Fig.~\ref{fig:potentials2}.

Compared to the previous cases of rotating `dust' shells, the consideration of $w\neq0$ brings about two notable new features. One is that we can now have true bounces in the presence of both interior and exterior horizons ---for the previous cases with $w=0$ true bounces were only allowed above the black dashed line, i.e., when there was no exterior horizon. Another remarkable difference is the existence of a novel dark-purple region (TB3) which corresponds to a true bounce satisfying the WEC but violating the DEC.

With a nonvanishing isotropic component of the pressure, there is the possibility of having, in addition to a centrifugal potential barrier at $\cR\sim (Ma^2)^{1/(2N+2)}$, also a pressure barrier at $\cR\sim m_0^{\frac{1}{(2N)}}E^{-\frac{1}{(2N+1)w}}$. [The position of these features in the radial potential can be straightforwardly inferred from Eq.~\eqref{eq:potential}.]
This property can be observed in Fig.~\ref{fig:potentials2}, especially in panel \textbf{H} where the two maxima of the potential are more evident. It is exactly this new pressure barrier that permits a true bounce outside the exterior horizon satisfying the DEC --- the centrifugal barrier occurs inside the horizon, and if we turned off the isotropic component of the pressure we would get a two-world orbit instead of a true bounce. The differences between panels \textbf{G} and \textbf{H} are only a consequence of changing the proper mass parameter $m_0$. For some intermediate value one would find a marginal orbit, a true bounce on the verge of becoming a two-world orbit. Such trajectories are associated with the nearly-vertical solid black line in the first panel of Fig.~\ref{fig:N1_varyw_varyE}, between the TWO1 and TB1 regions.

Fig.~\ref{fig:N1_varyw_varyE} allows us to infer some general behavior as the parameters $w$ and $E$ are varied. For the same $w$, increasing $E$ shifts the regions to the right (to larger values of $m_0$). The same effect is obtained by increasing $w$ for fixed $E$. Increasing the value of either $E$ or $w$ shrinks the domains where the WEC and the DEC are satisfied. For $E=0.25$ the DEC is satisfied in most part of the scanned region (dominated by TB1), but for $E=1$ the DEC is not satisfied anywhere, and the WEC is just satisfied in a small region (bottom-right corner).

Just like in the previous cases, we find no shell trajectories that would correspond to violations of cosmic censorship. Since we only plot the parameter space with $m_0 >0$ in Fig.~\ref{fig:N1_varyw_varyE}, the region where the interior geometry possesses a horizon but the exterior geometry is over-extremal is not even being shown.

\begin{figure}[!t]
\centering
\subfigure{\includegraphics[width=0.38\textwidth]{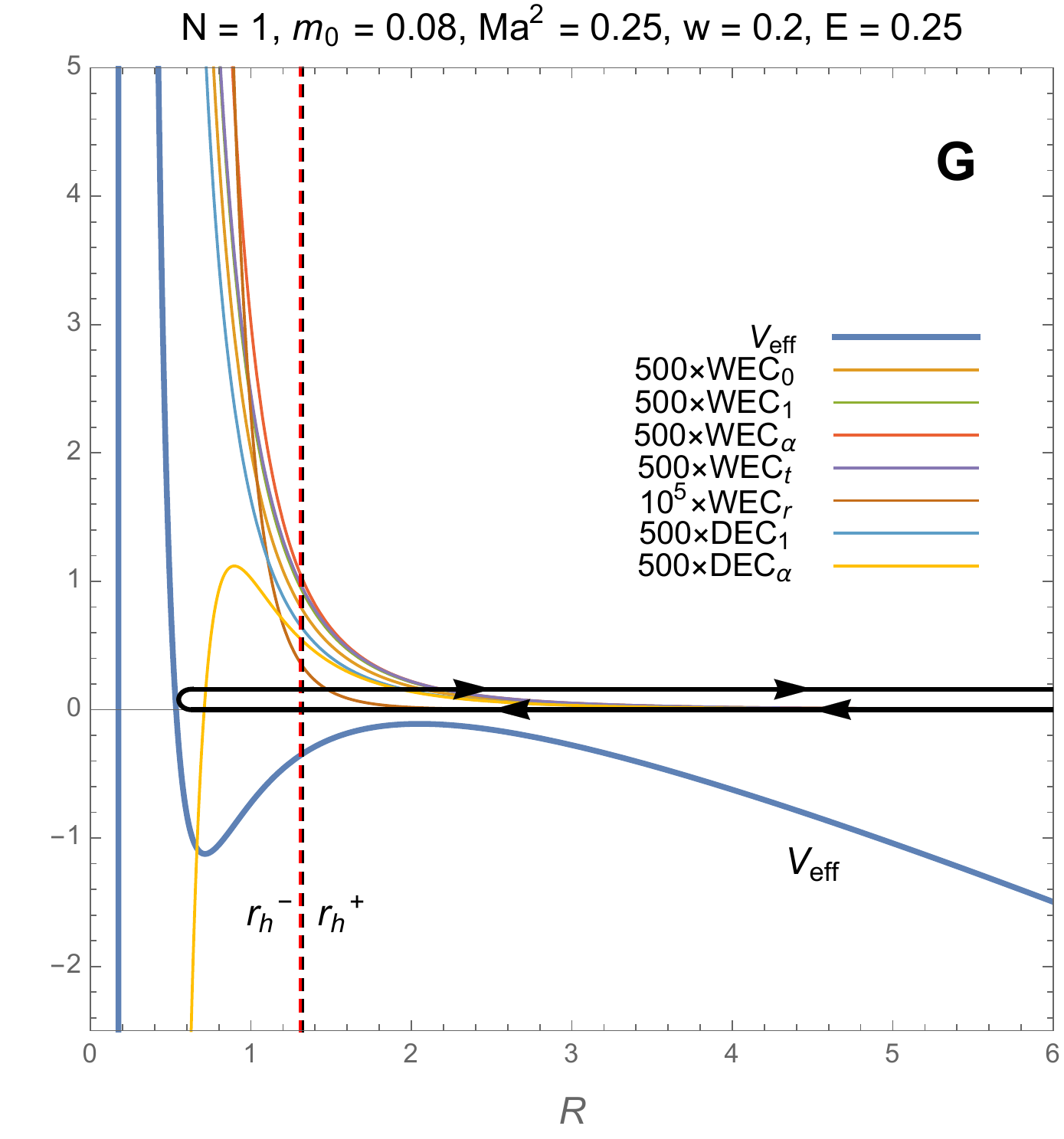}}
\quad
\subfigure{\includegraphics[width=0.38\textwidth]{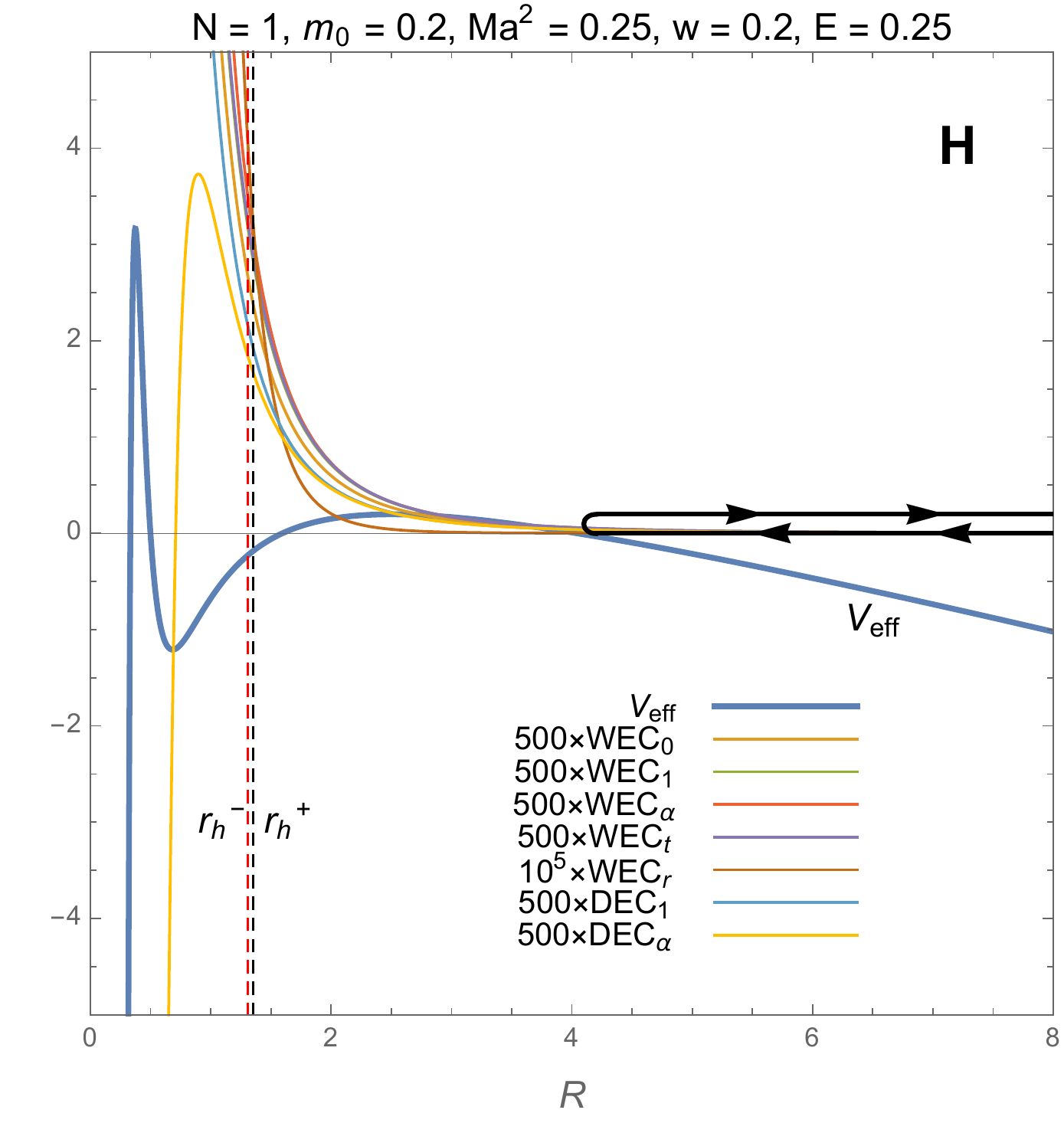}}
\\
\subfigure{\includegraphics[width=0.38\textwidth]{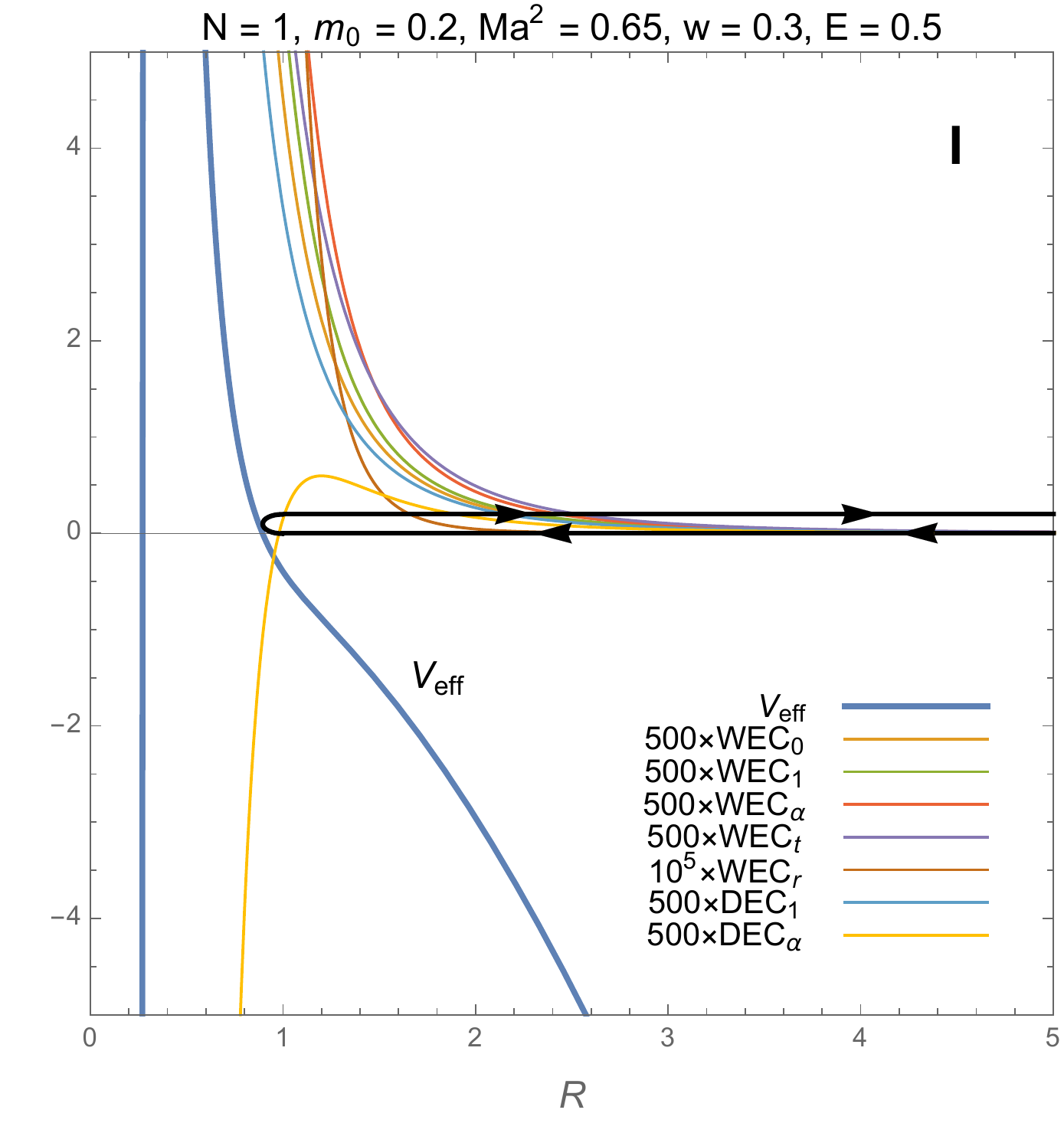}}
\quad
\subfigure{\includegraphics[width=0.38\textwidth]{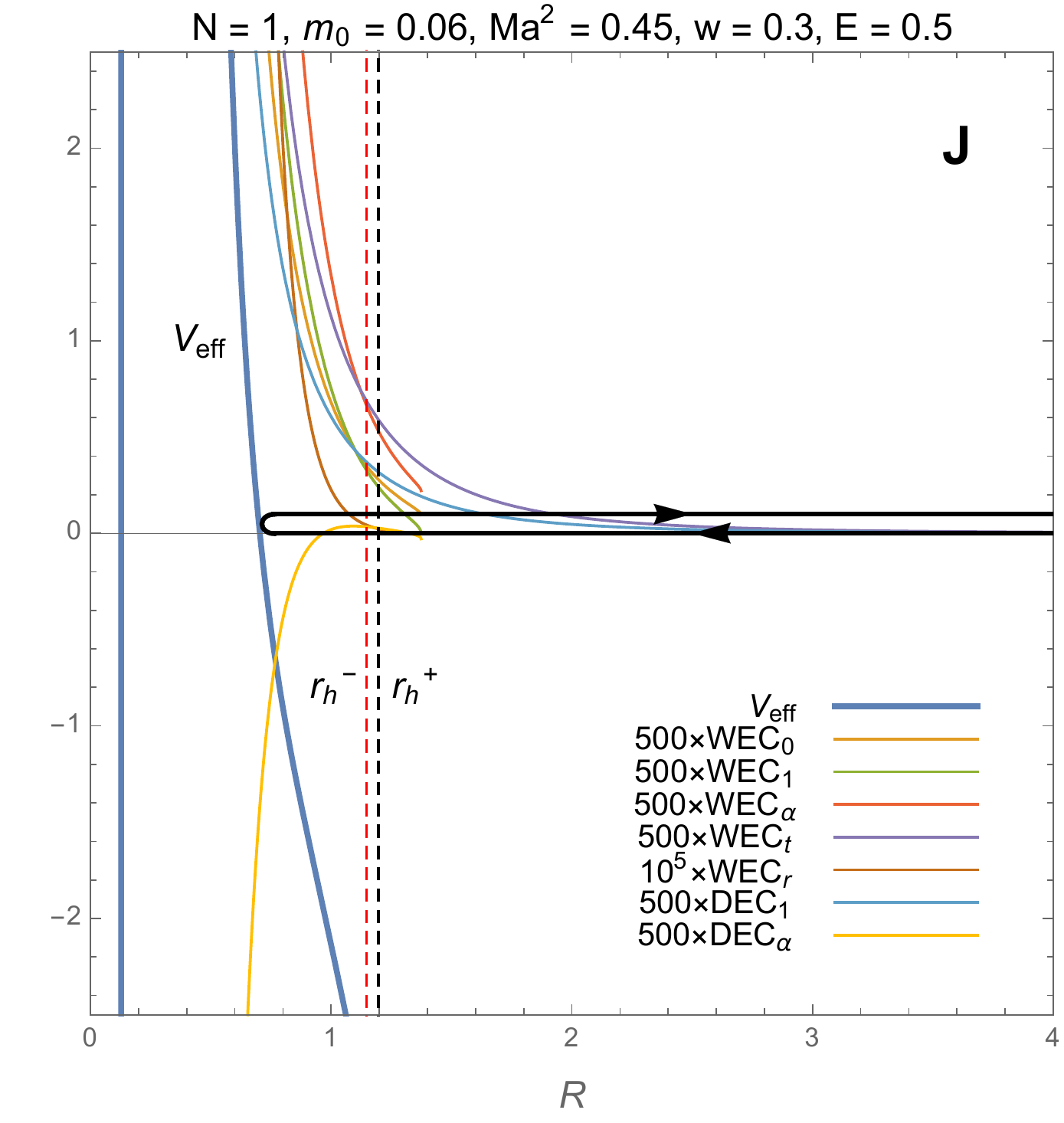}}
\caption{Plots of the radial potential $V_{eff}$ and the various energy conditions~\eqref{eq:WEC} and~\eqref{eq:DEC}, corresponding to choices of parameters indicated by the points \textbf{G--J} in the first and fourth panels of Fig.~\ref{fig:N1_varyw_varyE}. Panel \textbf{G} represents a two-world orbit satisfying the WEC but violating the DEC once inside the horizon. Panel \textbf{H} corresponds to a true bounce where both WEC and DEC are satisfied. Panel \textbf{I} represents a true bounce off a naked singularity, with WEC satisfied and DEC violated. Panel \textbf{J} illustrates a two-world orbit with WEC violated (and therefore also violating the DEC).}
\label{fig:potentials2}
\end{figure}

\section*{Acknowledgement}

We thank T\'erence Delsate for initial collaboration on this project.
We would also like to thank Masashi Kimura, Jos\'e Lemos, Jos\'e Nat\'ario, Alberto Saa and Jorge Santos for useful comments and correspondence.
JVR is grateful to {\it Universidade Federal de S\~ao Carlos} and {\it Universidade Estadual de Campinas}, for their kind hospitality.
JVR and RS would also like to thank FAPESP grants 2013/09357-9 and 2016/01343-7 for funding a visit to ICTP-SAIFR in May of 2017, where part of this work was done.
JVR acknowledges financial support from the European Union's Horizon 2020 research and innovation programme under the Marie Sk\l{}odowska-Curie grant agreement No REGMat-2014-656882 and under ERC Advanced Grant GravBHs-692951.
JVR was partially supported by the Spanish MINECO under Project No. FPA2013-46570-C2-2-P.

\bigskip

\appendix
\mysection{Bounds on parameters from energy conditions
\label{sec:boundsEC}}

In order to satisfy the energy conditions (null, weak or dominant) we need at the very least that the discriminant in Eqs.~\eqref{eq:lambda0} and~\eqref{eq:lambda1} be non-negative, i.e., $WEC_r\geq0$ must be obeyed.
Considering expressions~(\ref{eq:rho}--\ref{eq:DeltaP}), this imposes the following inequality:
\be
\left(m_0^{1+\frac{2N+1}{2N}w}\right)^2 \geq \frac{16(N+1)^2 Ma^2(\sqrt{M_+}-\sqrt{M_-})^2 \cR^{2(2N+1)w}}{(1+w)^2 \cR^2 \left(1+\frac{2Ma^2}{\cR^{2N+2}}\right)^{1-w} \left(1+2N+N\frac{2Ma^2}{\cR^{2N+2}}\right)^2}\,,
\label{eq:positive_discriminant}
\ee
whose compliance at large $\cR$ implies an upper bound on the EoS parameter,
\be
w\leq \frac{1}{2N+1}\,.
\ee

Assuming this holds, a further constraint can be inferred by studying the large $\cR$ behavior of the remaining conditions~\eqref{eq:WEC} and~\eqref{eq:DEC}. One finds that $WEC_1$ and $DEC_\alpha$ are automatically positive as $\cR\to\infty$, while $WEC_0$, $WEC_\alpha$, $WEC_t$ and $DEC_1$ are positive as $\cR\to\infty$ if and only if
\be
m_0^{1+\frac{2N+1}{2N}w} \geq 0\,.
\ee

The bounds obtained above are necessary conditions for the WEC and DEC to be satisfied.
Particularizing to the case of shells starting from rest at infinity ($w=0$ and $E=1$) and collapsing onto black holes ---as opposed to naked singularities--- we can give sufficient conditions for the validity of the WEC. Consider first the constraint $WEC_r \geq 0$. Inspection of~\eqref{eq:positive_discriminant} shows that it is satisfied at both large and small $\cR$. The question is whether $WEC_r \geq 0$ for \emph{all} values of $\cR$, which a fully collapsing shell will necessarily explore. For $w=0$ we can write~\eqref{eq:positive_discriminant} as
\be
\frac{m_0^2}{(\sqrt{M_+}-\sqrt{M_-})^2} \geq \frac{16(N+1)^2 Ma^2}{\cR^2 \left(1+\frac{2Ma^2}{\cR^{2N+2}}\right) \left(1+2N+N\frac{2Ma^2}{\cR^{2N+2}}\right)^2} \equiv \Lambda(\cR,Ma^2,N)\,. 
\label{eq:LambdaBound}
\ee
It can be easily shown that $\Lambda$, as a function of $\cR$, has a maximum value of the form
\be
\Lambda_{max}(Ma^2,N) = c_N(Ma^2)^{\frac{N}{N+1}}\,,
\ee
where $c_N$ is a dimension-dependent constant that satisfies $1<c_N<4$. Requiring the interior geometry to have a horizon covering the singularity translates into an upper bound on the spin:
\be
Ma^2\leq (Ma^2)_{max}\equiv \left[\frac{N}{N+1}\left(\frac{2}{N}\right)^{1/(N+1)} M_-\right]^{\frac{N+1}{N}}\,.
\ee
Thus, we see that the right hand side of~\eqref{eq:LambdaBound} is bounded from above by
\be
\Lambda(\cR,Ma^2,N) \leq \Lambda_{max}(Ma^2,N) < 4 M_-\,.
\ee
On the other hand, since we are assuming $E=1$ the left hand side is bounded from below:
\be
\frac{m_0^2}{(\sqrt{M_+}-\sqrt{M_-})^2}  = (\sqrt{M_+}+\sqrt{M_-})^2 \geq (2\sqrt{M_-})^2 = 4 M_-\,,
\ee
Thus, inequality~\eqref{eq:positive_discriminant} is satisfied for all $\cR$ when $w=0$ and $E=1$.

Moreover, given expressions~(\ref{eq:rho}--\ref{eq:DeltaP}) and~\eqref{eq:deltabeta}, the positivity of $WEC_0, WEC_1, WEC_\alpha$ and $WEC_t$ follows straightforwardly if $m_0\geq0$; otherwise these conditions are violated. (The same thing holds for $DEC_1$.)
In conclusion, for the case $w=0$, starting from rest at infinity and excluding over-extremal interior geometries, the shell's matter respects the weak energy condition throughout its entire motion if and only if $m_0\geq0$.

However, the status of the dominant energy condition differs from this. In particular, the condition
\be
DEC_\alpha = \frac{\rho-3P}{2}-\Delta P + \sqrt{\left(\frac{\rho+P}{2}\right)^2-\varphi^2} \geq 0
\ee
is necessarily violated at small enough $\cR$, where the pressure anisotropy term dominates. Note that Ref.~\cite{Nakao:2014qva} suggested that arbitrarily small over-spinning objects can exist, at least as transients, but within our construction it does not seem possible to obtain arbitrarily small over-spinning dust shells satisfying the dominant energy condition.

Nevertheless, if the exterior geometry has an event horizon (and $m_0\geq0$), then the radius at which the dominant energy condition is violated is always inside the horizon. This can be shown by noting that the radius $\cR_v$, determined by $DEC_\alpha|_{\cR=\cR_v}=0$, or equivalently
\be
\left(1+\frac{2Ma^2}{\cR_v^{2N+2}}\right) \left(1+2N-N\frac{2Ma^2}{\cR_v^{2N+2}}\right) = 2(N+1) \frac{\cR_v^{2N}}{m_0} \frac{(\sqrt{M_+}-\sqrt{M_-})^2}{m_0}\,,
\ee
is smaller than the radius of the exterior horizon, $r_h^+$, which obeys
\be
1-\frac{2M_+}{(r_h^+)^{2N}}+\frac{2Ma^2}{(r_h^+)^{2N+2}}=0\,.
\ee
Indeed, it is not hard to see that $DEC_\alpha|_{\cR=r_h^+}\geq0$ and, by continuity, this implies $\cR_v<r_h^+$.

\begin{figure}[t]
\centering
\subfigure{\includegraphics[width=0.38\textwidth]{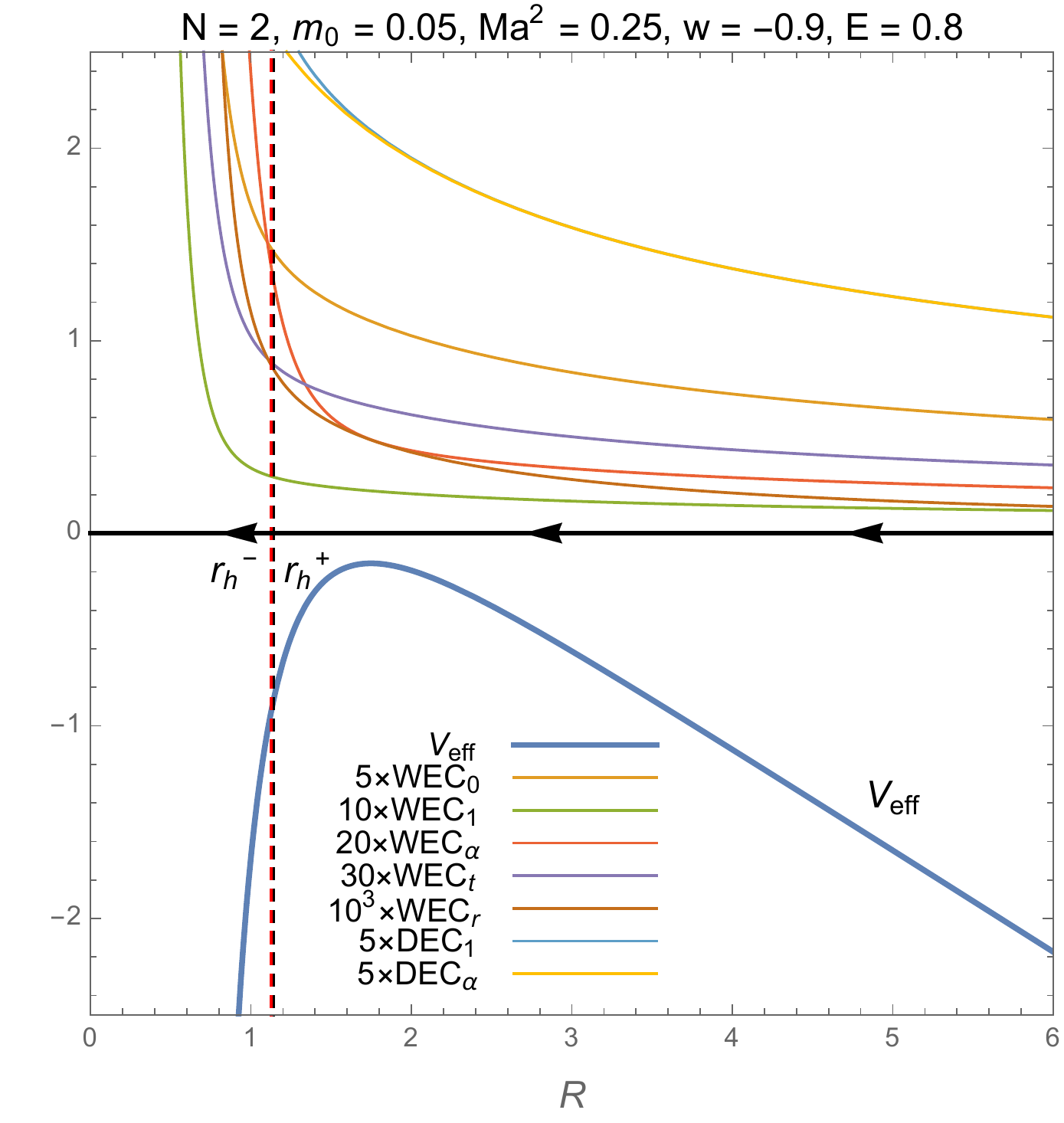}}
\quad
\subfigure{\includegraphics[width=0.38\textwidth]{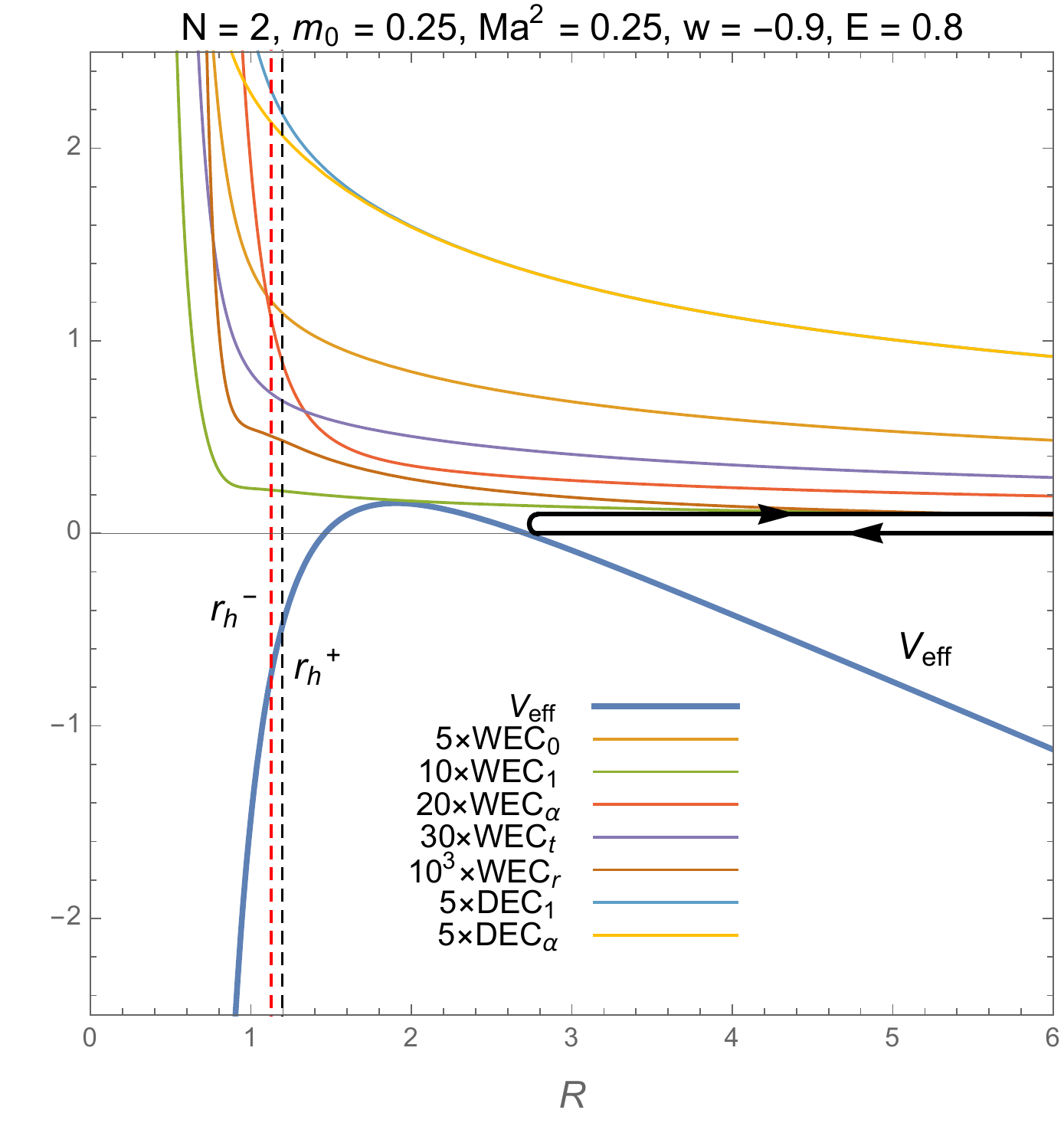}}
\caption{Examples of a full plunge (left) and a true bounce (right), satisfying both the weak and dominant energy conditions, for $N=2$ (i.e., a seven-dimensional spacetime) and with $w=-0.9$. This value was chosen as representative within the narrow interval~\eqref{eq:wnarrow} corresponding to tense (negative $w$) shells that are classically allowed to come in from infinity and that satisfy the DEC at arbitrarily small radii.}
\label{fig:FullPlunge_N2_wneg}
\end{figure}

There is a possibility that the dominant energy condition is satisfied even for full plunges, but this requires a negative isotropic pressure. Assuming that $WEC_r\geq0$ is satisfied ---so that conditions~\eqref{eq:WEC} and~\eqref{eq:DEC} are real--- the term in $DEC_\alpha$ that dominates at small radii is
\be
\left.DEC_\alpha\right|_{\cR\to0} \propto -(1+Nw)\cR^{-2-(3+w)N}\,.
\ee
Thus, for $w<-1/N$ there is a chance that the DEC is satisfied everywhere. This is indeed the case, as shown explicitly in Fig.~\ref{fig:FullPlunge_N2_wneg}. However, since the condition $WEC_t\geq0$ is equivalent to $w\geq-1$, it is clear that this scenario requires $N\geq2$.

\mysection{Polytropic equation-of-state
\label{sec:poly}}

Polytropic equations-of-state refer to non-linear relations between the pressure and energy density of the following form:
\be
P=w\rho\left(\frac{\rho}{\gamma}\right)^{1/n}\,,
\ee
where $n$ is known as the polytropic index. In the limit $n\to\infty$ one recovers the linear EoS considered in the main text.

Inserting expressions~(\ref{eq:rho}) and (\ref{eq:P}) into this relation we obtain
\be
\frac{d}{d\cR}\left[\left(-\cR^{2N}[[\beta(\cR)]]\right)^{-1/n}\right] = 
\frac{w}{n(8\pi \gamma)^{1/n}} \frac{\left(\frac{d}{d\cR}[\cR^{2N}h(\cR)]\right)^{1+1/n}}{\cR^{2N+\frac{4N+1}{n}}h(\cR)}\,.
\ee
Upon integration, this gives
\be
[[\beta(\cR)]]= -\frac{8\pi \gamma}{\cR^{2N}} \left(\frac{n}{w\, {\cal I}(\cR)}\right)^n\,,
\ee
where
\be
{\cal I}(\cR) = \int d\cR \frac{\left(\frac{d}{d\cR}[\cR^{2N}h(\cR)]\right)^{1+1/n}}{\cR^{2N+\frac{4N+1}{n}}h(\cR)}\,.
\label{eq:integral}
\ee
It turns out the integral~\eqref{eq:integral} can be computed in terms of hypergeometric functions. The result is
\begin{flalign}
{\cal I}(\cR) &= \frac{n\, \cR^{-\frac{4N+1}{n}}}{3N+2} \left(\frac{2N Ma^2}{\cR\, h(\cR)}\right)^{\frac{1}{n}} \left(1+\frac{\cR^{2N+2}}{2Ma^2}\right)^{\frac{1}{2n}} \nonumber\\ 
& \!\!\!\!\!\! \times \left[(N+1) F_1\left(-\frac{3N+2}{(2N+2) n};1+\frac{1}{2n},-\frac{1}{n};1-\frac{3N+2}{(2N+2) n};-\frac{\cR^{2N+2}}{2Ma^2},-\frac{(2N+1) \cR^{2N+2}}{2N Ma^2}\right)\right. \nonumber\\
& \;\; \left.-(2N+1) F_1\left(-\frac{3N+2}{(2N+2) n};\frac{1}{2n},-\frac{1}{n};1-\frac{3N+2}{(2N+2) n};-\frac{\cR^{2N+2}}{2Ma^2},-\frac{(2N+1) \cR^{2N+2}}{2N Ma^2}\right)\right],
\end{flalign}
where $F_1(a;b,c;d; x,y)$ denotes the two-variable Appell hypergeometric function.

Analogously to the case of a linear equation-of-state, we can again obtain the radial equation of motion for the shell in the form~\eqref{eq:motion}, but now the effective potential is given by
\beq
V_{\rm eff}(\cR) &=& \frac{1}{2}\left(g_+^{-2}(\cR)+g_-^{-2}(\cR)\right) - \frac{h(\cR)^2}{4\cR^2} \left(\frac{8\pi \gamma}{\cR^{2N}} \left(\frac{n}{w\, {\cal I}(\cR)}\right)^n\right)^2 - \frac{\cR^2}{4h(\cR)^2} \frac{\left(g_+^{-2}(\cR)-g_-^{-2}(\cR)\right)^2}{\left(\frac{8\pi \gamma}{\cR^{2N}} \left(\frac{n}{w\, {\cal I}(\cR)}\right)^n\right)^2} \nonumber \\
&=& 1-\frac{M_-+M_+}{\cR^{2N}}+\frac{2Ma^2}{\cR^{2N+2}}- \left(1+\frac{2Ma^2}{\cR^{2N+2}}\right) \left(\frac{4\pi \gamma}{\cR^{2N}}\right)^2 \left(\frac{n}{w\, {\cal I}(\cR)}\right)^{2n} \nonumber\\
&& - \left(1+\frac{2Ma^2}{\cR^{2N+2}}\right)^{-1} \left(\frac{8\pi \gamma}{\cR^{2N}}\right)^{-2} \left(\frac{n}{w\, {\cal I}(\cR)}\right)^{-2n} \left(\frac{M_+-M_-}{\cR^{2N}}\right)^2\,.
\label{eq:potential2}
\eeq
Having determined the radial potential, one can proceed to study the possible shell trajectories as the parameters are varied, along the lines of Section~\ref{sec:scan}. Such an analysis is beyond the scope of this paper and merits a separate study on its own.

\bigskip


\begin{thebibliography}{99}


\bibitem{Oppenheimer:1939ue} 
  J.~R.~Oppenheimer and H.~Snyder,
  ``On Continued gravitational contraction,''
  Phys.\ Rev.\  {\bf 56}, 455 (1939).

\bibitem{Nakamura:1981} 
  T.~Nakamura,
  ``General Relativistic Collapse of Axially Symmetric Stars Leading to the Formation of Rotating Black Holes,''
  Prog.\ Theor.\ Phys.\ {\bf 65}, 1876 (1981).

\bibitem{Stark:1985da} 
  R.~F.~Stark and T.~Piran,
  ``Gravitational Wave Emission From Rotating Gravitational Collapse,''
  Phys.\ Rev.\ Lett.\  {\bf 55}, 891 (1985)
  Erratum: [Phys.\ Rev.\ Lett.\  {\bf 56}, 97 (1986)].



\bibitem{Penrose:1969pc}
  R.~Penrose,
  ``Gravitational collapse: The role of general relativity,''
  Riv.\ Nuovo Cimento  {\bf 1}, 252 (1969) [Gen.\ Relativ.\ Gravit.\ {\bf 34}, 1141 (2002)].



\bibitem{Shibata:2000gt} 
  M.~Shibata,
  ``Axisymmetric simulations of rotating stellar collapse in full general relativity: Criteria for prompt collapse to black holes,''
  Prog.\ Theor.\ Phys.\  {\bf 104}, 325 (2000)
  [gr-qc/0007049].



\bibitem{Kerr:1963} 
  R.~P.~Kerr,
  ``Gravitational field of a spinning mass as an example of algebraically special metrics,''
  Phys.\ Rev.\ Lett.\  {\bf 11}, 237 (1963).



\bibitem{Abrahams:1994ge} 
  A.~M.~Abrahams, G.~B.~Cook, S.~L.~Shapiro and S.~A.~Teukolsky,
  ``Solving Einstein's equations for rotating space-times: Evolution of relativistic star clusters,''
  Phys.\ Rev.\ D {\bf 49}, 5153 (1994).

\bibitem{Giacomazzo:2011cv} 
  B.~Giacomazzo, L.~Rezzolla and N.~Stergioulas,
  ``Collapse of differentially rotating neutron stars and cosmic censorship,''
  Phys.\ Rev.\ D {\bf 84}, 024022 (2011)
  [arXiv:1105.0122 [gr-qc]].

\bibitem{Sperhake:2009jz} 
  U.~Sperhake, V.~Cardoso, F.~Pretorius, E.~Berti, T.~Hinderer and N.~Yunes,
  ``Cross section, final spin and zoom-whirl behavior in high-energy black hole collisions,''
  Phys.\ Rev.\ Lett.\  {\bf 103}, 131102 (2009)
  [arXiv:0907.1252 [gr-qc]].



\bibitem{Wald:1974} 
  R.~M.~Wald,
  ``Gedanken Experiments to Destroy a Black Hole,''
  Annals Phys.\ {\bf 83}, 548 (1974).

\bibitem{BouhmadiLopez:2010vc} 
  M.~Bouhmadi-L\'opez, V.~Cardoso, A.~Nerozzi and J.~V.~Rocha,
  ``Black holes die hard: can one spin-up a black hole past extremality?,''
  Phys.\ Rev.\ D {\bf 81}, 084051 (2010)
  [arXiv:1003.4295 [gr-qc]].

\bibitem{Rocha:2014gza} 
  J.~V.~Rocha, R.~Santarelli and T.~Delsate,
  ``Collapsing rotating shells in Myers-Perry-AdS$_5$ spacetime: A perturbative approach,''
  Phys.\ Rev.\ D {\bf 89}, 104006 (2014)
  [arXiv:1402.4161 [gr-qc]].

\bibitem{Rocha:2014jma} 
  J.~V.~Rocha and R.~Santarelli,
  ``Flowing along the edge: spinning up black holes in AdS spacetimes with test particles,''
  Phys.\ Rev.\ D {\bf 89}, no. 6, 064065 (2014)
  [arXiv:1402.4840 [gr-qc]].

\bibitem{Natario:2016bay} 
  J.~Nat\'ario, L.~Queimada and R.~Vicente,
  ``Test fields cannot destroy extremal black holes,''
  Class.\ Quant.\ Grav.\  {\bf 33}, no. 17, 175002 (2016)
  [arXiv:1601.06809 [gr-qc]].

\bibitem{Jacobson:2009kt} 
  T.~Jacobson and T.~P.~Sotiriou,
  ``Over-spinning a black hole with a test body,''
  Phys.\ Rev.\ Lett.\  {\bf 103}, 141101 (2009)
  Erratum: [Phys.\ Rev.\ Lett.\  {\bf 103}, 209903 (2009)]
  [arXiv:0907.4146 [gr-qc]].


\bibitem{Barausse:2010ka} 
  E.~Barausse, V.~Cardoso and G.~Khanna,
  ``Test bodies and naked singularities: Is the self-force the cosmic censor?,''
  Phys.\ Rev.\ Lett.\  {\bf 105}, 261102 (2010)
  [arXiv:1008.5159 [gr-qc]].

\bibitem{Sorce:2017dst} 
  J.~Sorce and R.~M.~Wald,
  ``Gedanken Experiments to Destroy a Black Hole II: Kerr-Newman Black Holes Cannot be Over-Charged or Over-Spun,''
  Phys.\ Rev.\ D {\bf 96}, no. 10, 104014 (2017)
  [arXiv:1707.05862 [gr-qc]].




\bibitem{Lehner:2010pn} 
  L.~Lehner and F.~Pretorius,
  ``Black Strings, Low Viscosity Fluids, and Violation of Cosmic Censorship,''
  Phys.\ Rev.\ Lett.\  {\bf 105}, 101102 (2010)
  [arXiv:1006.5960 [hep-th]].

\bibitem{Figueras:2015hkb} 
  P.~Figueras, M.~Kunesch and S.~Tunyasuvunakool,
  ``End Point of Black Ring Instabilities and the Weak Cosmic Censorship Conjecture,''
  Phys.\ Rev.\ Lett.\  {\bf 116}, no. 7, 071102 (2016)
  [arXiv:1512.04532 [hep-th]].

\bibitem{Figueras:2017zwa} 
  P.~Figueras, M.~Kunesch, L.~Lehner and S.~Tunyasuvunakool,
  ``End Point of the Ultraspinning Instability and Violation of Cosmic Censorship,''
  Phys.\ Rev.\ Lett.\  {\bf 118}, no. 15, 151103 (2017)
  [arXiv:1702.01755 [hep-th]].



\bibitem{Cohen:1968} 
  J.~M.~Cohen,
  ``Gravitational Collapse of Rotating Bodies,''
  Phys.\ Rev. {\bf 173}, 1258 (1968).

\bibitem{Lindblom:1974bq} 
  L.~Lindblom and D.~R.~Brill,
  ``Inertial effects in the gravitational collapse of a rotating shell,''
  Phys.\ Rev.\ D {\bf 10}, 3151 (1974).

\bibitem{Wagoner:1965} 
  R.~V.~Wagoner,
  ``Rotation and Gravitational Collapse,''
  Phys.\ Rev. {\bf 138}, B1583 (1965).




\bibitem{Crisostomo:2003xz} 
  J.~Cris\'ostomo and R.~Olea,
  ``Hamiltonian treatment of the gravitational collapse of thin shells,''
  Phys.\ Rev.\ D {\bf 69}, 104023 (2004)
  [hep-th/0311054].

\bibitem{Mann:2008rx} 
  R.~B.~Mann, J.~J.~Oh and M.~-I.~Park,
  ``The Role of Angular Momentum and Cosmic Censorship in the (2+1)-Dimensional 
Rotating Shell Collapse,''
  Phys.\ Rev.\ D {\bf 79}, 064005 (2009)
  [arXiv:0812.2297 [hep-th]].

\bibitem{Vaz:2008uv} 
  C.~Vaz and K.~R.~Koehler,
  ``A Rotating, Inhomogeneous Dust Interior for the BTZ Black Hole,''
  Phys.\ Rev.\ D {\bf 78}, 024038 (2008)
  [arXiv:0805.1908 [gr-qc]].



\bibitem{Delsate:2014iia} 
  T.~Delsate, J.~V.~Rocha and R.~Santarelli,
  ``Collapsing thin shells with rotation,''
  Phys.\ Rev.\ D {\bf 89}, 121501(R) (2014)
  [arXiv:1405.1433 [gr-qc]].

\bibitem{Rocha:2015tda} 
  J.~V.~Rocha,
  ``Gravitational collapse with rotating thin shells and cosmic censorship,''
  Int.\ J.\ Mod.\ Phys.\ D {\bf 24}, no. 09, 1542002 (2015)
  [arXiv:1501.06724 [gr-qc]].

\bibitem{Bizon:2005cp} 
  P.~Bizo\'n, T.~Chmaj and B.~G.~Schmidt,
  ``Critical behavior in vacuum gravitational collapse in 4+1 dimensions,''
  Phys.\ Rev.\ Lett.\  {\bf 95}, 071102 (2005)
  [gr-qc/0506074].

\bibitem{Natario:2017szw}
  J.~Nat\'ario, L.~Queimada and R.~Vicente,
  ``Rotating elastic string loops in flat and black hole spacetimes: stability, cosmic censorship and the Penrose process,''
  Class.\ Quant.\ Grav.\  {\bf 35}, no. 7, 075003 (2018)
  [arXiv:1712.05416 [gr-qc]].


\bibitem{Poisson:2004} 
  E.~Poisson,
  ``A Relativist's Toolkit: The Mathematics of Black-Hole Mechanics,''
  Cambridge University Press (2004).
 
 

\bibitem{Myers:1986un}
  R.~C.~Myers and M.~J.~Perry,
  ``Black Holes In Higher Dimensional Space-Times,''
  Annals Phys.\  {\bf 172}, 304 (1986).



\bibitem{Shibata:2008rq} 
  M.~Shibata, H.~Okawa and T.~Yamamoto,
  ``High-velocity collision of two black holes,''
  Phys.\ Rev.\ D {\bf 78}, 101501 (2008)
  [arXiv:0810.4735 [gr-qc]].

\bibitem{Pollney:2009yz} 
  D.~Pollney, C.~Reisswig, E.~Schnetter, N.~Dorband and P.~Diener,
  ``High accuracy binary black hole simulations with an extended wave zone,''
  Phys.\ Rev.\ D {\bf 83}, 044045 (2011)
  [arXiv:0910.3803 [gr-qc]].



%
%



\bibitem{Frolov:2002xf} 
  V.~P.~Frolov and D.~Stojkovic,
  ``Quantum radiation from a five-dimensional rotating black hole,''
  Phys.\ Rev.\ D {\bf 67}, 084004 (2003)
  [gr-qc/0211055].

\bibitem{Kunduri:2006qa} 
  H.~K.~Kunduri, J.~Lucietti and H.~S.~Reall,
  ``Gravitational perturbations of higher dimensional rotating black holes:
Tensor perturbations,''
  Phys.\ Rev.\ D {\bf 74}, 084021 (2006)
  [hep-th/0606076].

\bibitem{Hoxha:2000jf} 
  P.~Hoxha, R.~R.~Martinez-Acosta and C.~N.~Pope,
  ``Kaluza-Klein consistency, Killing vectors, and Kahler spaces,''
  Class.\ Quant.\ Grav.\  {\bf 17}, 4207 (2000)
  [hep-th/0005172].

\bibitem{Dias:2010eu} 
  O.~J.~C.~Dias, P.~Figueras, R.~Monteiro, H.~S.~Reall and J.~E.~Santos,
  ``An instability of higher-dimensional rotating black holes,''
  JHEP {\bf 1005}, 076 (2010)
  [arXiv:1001.4527 [hep-th]].



\bibitem{Israel:1966rt} 
  W.~Israel,
  ``Singular hypersurfaces and thin shells in general relativity,''
  Nuovo Cim.\ B {\bf 44S10}, 1 (1966)
  [Erratum-ibid.\ B {\bf 48}, 463 (1967)]
  [Nuovo Cim.\ B {\bf 44}, 1 (1966)].

  \bibitem{Darmois}
  G.~Darmois,
  ``Les \'equations de la gravitation einsteinienne'', Chapitre V,
  M\'emorial de Sciences Math\'ematiques, fascicule XXV (1927).



\bibitem{Gao:2008jy} 
  S.~Gao and J.~P.~S.~Lemos,
  ``Collapsing and static thin massive charged dust shells in a Reissner-Nordstrom black hole background in higher dimensions,''
  Int.\ J.\ Mod.\ Phys.\ A {\bf 23}, 2943 (2008)
  [arXiv:0804.0295 [hep-th]].



\bibitem{Wald:1984rg} 
  R.~M.~Wald,
  ``General Relativity,''
  Chicago, USA: University Press (1984).

\bibitem{Kolassis:1988}
  C.~A.~Kolassis, N.~O.~Santos and D.~Tsoubelis,
  ``Energy conditions for an imperfect fluid,''
  Class.\ Quant.\ Grav.\ {\bf 5} 1329 (1988).

\bibitem{Kuchar:1990vy} 
  K.~V.~Kucha\v{r} and C.~G.~Torre,
  ``Gaussian reference fluid and interpretation of quantum geometrodynamics,''
  Phys.\ Rev.\ D {\bf 43}, 419 (1991).



\bibitem{Diemer:2014lba} 
  V.~Diemer, J.~Kunz, C.~L\"ammerzahl and S.~Reimers,
  ``Dynamics of test particles in the general five-dimensional Myers-Perry spacetime,''
  Phys.\ Rev.\ D {\bf 89}, no. 12, 124026 (2014)
  [arXiv:1404.3865 [gr-qc]].



\bibitem{Dias:2011jg} 
  O.~J.~C.~Dias, R.~Monteiro and J.~E.~Santos,
  ``Ultraspinning instability: the missing link,''
  JHEP {\bf 1108}, 139 (2011)
  [arXiv:1106.4554 [hep-th]].



\bibitem{Nakao:2014qva} 
  K.~i.~Nakao, M.~Kimura, T.~Harada, M.~Patil and P.~S.~Joshi,
  ``How small can an over-spinning body be in general relativity?,''
  Phys.\ Rev.\ D {\bf 90}, no. 12, 124079 (2014)
  [arXiv:1406.6798 [gr-qc]].















\end{thebibliography}
\end{document}